\documentclass[useAMS,usenatbib]{mnras}
\usepackage{amsmath}

\usepackage{graphicx}
\usepackage{multirow} 

\usepackage{amssymb}
\newcommand\NS{N\!S}

\title[Tidally induced Crustal Failure on Icy Moons]{Crustal Failure on Icy Moons  from a Strong Tidal Encounter}
\author[Quillen et al.]{ Alice C. Quillen$^{1,2}$, David Giannella$^1$, John G. Shaw$^3$, 
\& Cynthia Ebinger$^4$
\\
$^1$Department of Physics and Astronomy, University of Rochester, Rochester, NY 14627, USA \\
$^2$ \texttt{aquillen@pas.rochester.edu}\\
$^3$Laboratory for Laser Energetics,  250 East River Rd.,  Rochester, NY 14623, USA \\
$^4$Department of Earth and Environmental Sciences, University of Rochester, Rochester, NY 14627, USA\\
}

\begin{document}

\maketitle

\begin{abstract}
 Close tidal encounters among large planetesimals and moons should have been more common than grazing  or normal impacts.
Using a mass spring model within an N-body simulation, we simulate the deformation of the surface of an elastic spherical body
caused by a close parabolic tidal encounter with a  body that has similar mass as that of the primary body.
Such an encounter can  induce sufficient stress on the surface to cause brittle failure of an icy crust and simulated fractures  can extend a large fraction of the radius of body.   Strong tidal encounters may be responsible for   
the formation of long graben complexes and chasmata  in ancient terrain of icy moons  such as Dione, Tethys, Ariel and Charon.
\end{abstract}


\begin{equation} \nonumber
\end{equation}

\section{Introduction}

A number of large surface 
features in the solar system have origins potentially due to giant impacts that occurred between planet-sized bodies
or planets and large planetesimals.  These include the crustal dichotomy of Mars \citep{wilhelms84,frey88,marinova08}
and the Moon \citep{jutzi11},  
jumbled terrain on Mercury \citep{schultz75} and numerous impact craters throughout the solar system
\citep{melosh89}.
Because of their larger cross-section, grazing impacts among planets and planetesimals
are more likely than normal angle impacts \citep{asphaug10}.
Likewise,  strong {\it tidal encounters}, 
those involving close encounters between two large bodies that do not actually touch, 
are more likely than grazing impacts.     
A fraction of
a body's gravitational binding energy can be dissipated during a close tidal encounter \citep{press77}.   
While some large surface features on planets and moons have proposed origins due to large
impacts, so far none have been linked to single close and strong tidal encounters. 

Strong tidal encounters are unlikely now, but would have occurred in the past,
during the late-heavy bombardment era and beforehand.
Orbits of closely packed moons can become unstable (e.g., \citealt{french12,cheng14}) 
and this too could cause close encounters between similar mass bodies.
Impacts primarily  cause compressive stress  \citep{melosh89}, but 
tidal stress can be tensile and many materials are weaker when subjected to tensile stress than a
comparable magnitude of compressive stress (for ice see Figure 1 by \citealt{petrovic03} and Figure 7.2 
by \citealt{collins10} and for the Earth's lithosphere
see failure envelopes in Figure 6.24 and Yield Strength envelopes in Figure 6.27 by \citealt{watts01} 
or Figure 9.6 by \citealt{kohlstedt10}).
Ancient regions of  planets and moons exhibit 
features such as chasmata, grooves, grabens
or graben complexes that are associated with extension and tensile deformation \citep{collins10}, and some of these
may have been caused by strong tidal encounters with large bodies. 
	
Many studies of tidal encounters between two planetesimals or between a planetesimal and a planet 
have  focused on tidal disruption (e.g., \citealt{dobro90,boss94,richardson98,sharma06,holsapple08}). 
But tidal stresses due to close encounters between
bodies can affect body rotation and shape \citep{bottke99} and
disturb weathered surfaces of asteroids,
exposing fresh surface materials \citep{binzel10,nesvorny10}. 
Simulations of granular materials have  predicted 
resurfacing in weak regions where  tidal stresses cause  avalanches or landslides \citep{yu14}.

Some icy moons exhibit  global tectonic features, such as grooves or long fractures,  that could be caused by varying tidal stresses 
exerted by their host planet (e.g., \citealt{helfenstein85,mcewen86,hurford07,smithkontor08,wahr09,hurford15}). 
The patterns and individual morphologies of parallel sets of grooves and  troughs
on satellites and asteroids such as Phobos, Eros, Ida, Gaspra, Epimetheus and Pandora (see \citealt{thomas10} for a review)
can be attributed to fracturing in weak materials caused by oscillating tidal stresses associated with orbital eccentricity
 \citep{morrison09} or an increase in tidal stress resulting from the orbital decay of the body itself \citep{soter77,hurford15}.
 For Phobos, the length of the grooves is perpendicular to the oscillating tensile stress \citep{morrison09}.
Long (130 km) linear fractures termed
``tiger stripes'' on Enceladus are connected to diurnal tidal stress variations \citep{smithkontor08,nimmo07,hurford07}.

Dione, Tethys, Rhea (moons of Saturn),  Titania (moon of Uranus) and Charon (moon of Pluto)
 have  heavily cratered surfaces and also display long faults extending
a significant fraction of the moon radius, and chasma and faults in pairs interpreted as graben or graben complexes
(for a review see section 6.4 by \citealt{collins10} and for recent results on the Pluto system see \citealt{stern15}).  
The Ithaca Chasma on Tethys is interpreted to be a large
graben complex \citep{giese07} and crater counts indicate that it is older than the large Odysseus impact basin 
(with radius 0.4 times that of the moon itself).
On Dione, Cassini imagery revealed that some fault networks have vertical offsets that dissect craters,  confirming their extensional tectonic origin and suggesting that they were formed early \citep{jaumann09}. 
These studies
suggest that the formation of chasmata and graben complexes can take place before or during an epoch of large impacts
and so during an epoch when strong tidal encounters would have occurred.
Explanations for the graben complexes include heating and expansion of the interior (e.g., \citealt{hillier91})
and stresses induced by reorientation following a large impact \citep{nimmo07}.
Strong tidal encounters have not yet been explored as a possible explanation for the formation of surface features
such as chasmata and graben complexes on icy moons.

In this study we focus on the intersection between the works introduced above.   We  consider rare and close
tidal encounters, that might have occurred billions of years ago, 
between large bodies that are not gravitationally bound (not in orbit
about each other).
The time of a parabolic or hyperbolic tidal encounter can be a few hours, so tidal encounters are extremely fast compared
to the time scales of most geophysical processes.   On such a short time scale rock and ice should  deform in a brittle-elastic mode
rather than a ductile, plastic or visco-elastic mode (e.g., \citealt{turcotte02,burgmann08}).  
To numerically simulate  tidal deformation we use a mass-spring model to simulate both elastic
response and gravity (see \citealt{tide}). Brittle failure is modeled by allowing springs on the surface to fail if they exceed
a critical tensile strain value.  
Our simulations allow to us to visualize brittle crustal failure following a hypothetical strong
tidal encounter.    Our approach
differs from the  granular flow simulations by \citet{schwartz13} with spring-like forces between
neighboring soft spheres that mimic cohesion and can simulate bulk tensile failure.

\subsection{Tidal encounters}

Following \citet{press77} (also see \citealt{ogilvie14}) the response of a body during a tidal encounter
can be estimated using an impulse approximation.
The maximum tidal force, $F_T$, on body $M$ from body $m$ during the encounter is approximately 
\begin{equation}
F_T \sim \frac{Gm R}{q^3} \end{equation}
where $q$ is the distance between body centers at closest approach (pericenter),  $m$ is the
mass of the tidal perturber, $R$ is the radius of the primary body with mass $M$, and $G$ is the gravitational
constant.
The time scale of the encounter is 
\begin{equation}
t_{enc} \sim 2 q/V_q \end{equation}
where $V_q$ is the velocity at pericenter. Together $F_T$ and $t_{enc}$  cause
a velocity perturbation on the surface of the primary body 
\begin{equation}
\Delta v \sim \frac{2 G m R}{q^2 V_q} \label{eqn:Deltav}
\end{equation}
If the orbit is parabolic then $\Delta v/\sqrt{GM/R} = \eta$, a dimensionless parameter
used to characterize parabolic tidal encounters \citep{press77},
that is
the ratio of acceleration due to self-gravity and the tidal acceleration at the body's surface. 

The extent of the tidal deformation of body $M$ can be estimated by balancing the kinetic energy per unit mass
due to the tidal impulse with elastic energy per unit mass
\begin{equation}  \Delta v^2 \sim \epsilon^2 E/\rho \end{equation}
with $E$ the Young's modulus and $\rho$ the density of body $M$, giving a strain of
\begin{equation}
 \epsilon \sim \left(\frac{e_g}{E}\right)^\frac{1}{2} \left( \frac{R}{q}\right)^{2} \left( \frac{m}{M} \right)
 \left({ \frac{v_c}{V_q} } \right), 
 \label{eqn:strain2}
\end{equation}
where 
\begin{equation} v_c =\sqrt{GM/R} \label{eqn:vc} 
\end{equation}
 is the velocity of a particle in a circular orbit grazing the surface of $M$ and 
 $e_g$
\begin{eqnarray}
e_g &\equiv & \frac{GM^2}{R^4} 
=  \left(\frac{4\pi}{3}\right)^2 GR^2 \rho^2  \nonumber \\
&=& 1.2  {\rm ~GPa} \left(\frac{R}{ 1000{\rm km}} \right)^2 \left(\frac{\rho }{1 {\rm g~cm}^3} \right)^2
\label{eqn:eg}
\end{eqnarray}
is approximately the gravitational binding energy density of body $M$ (and to order of magnitude its central pressure).
We have assumed that $M$ is a homogenous and spherical body and remains so during the encounter.


A time scale for elastic response can be estimated from the speed of elastic waves and the radius
of the primary body.
\begin{equation}
t_{elas} \sim \frac{R}{\sqrt{E/\rho}} = \left( \frac{e_g}{E} \right)^\frac{1}{2} t_{grav} \end{equation}
where we have defined a gravitational time scale
\begin{eqnarray}
t_{grav} &\equiv& \sqrt{ \frac{R^3}{GM} } = \sqrt{\frac{3}{4 \pi G \rho}} \nonumber \\
&\sim& 2000 {\rm s} \left( \frac{\rho}{1{\rm~g~cm}^{-3}} \right)^{-\frac{1}{2}}  \label{eqn:tgrav}
\end{eqnarray}
equivalent to the the inverse of the angular rotation rate of a  particle in a circular orbit 
grazing the surface of the body $M$.  The gravitational time scale  is only dependent on the body's mean density $\rho$.
For a Poisson modulus of $\nu=1/4$, the speed for P-waves is $V_P \approx 1.1 \sqrt{E/\rho}$
and that for S waves $V_S \approx 0.6 V_P$.
To order of magnitude $t_{elas}^{-1}$ is equal to the frequency of the slowest  vibrational mode of the body.

The strain rate (on $M$) during the tidal encounter can be estimated from the time scale for elastic response
\begin{eqnarray}
\dot \epsilon &\sim & \frac{\epsilon}{t_{elas}}  \nonumber \\
&\sim&  \left( \frac{R}{q}\right)^2 \left( \frac{m}{M} \right) \left({ \frac{v_c}{V_q} } \right) t_{grav}^{-1} \nonumber \\
&\sim&10^{-4}{\rm s}^{-1}
 \left( \frac{\rho}{1~{\rm g~cm}^{-3}} \right)^\frac{1}{2}  \left( \frac{R}{q}\right)^2 \left( \frac{m}{M} \right) \left({ \frac{v_c}{V_q} } \right)
\end{eqnarray} 
This expression implies that close encounters between similar mass bodies are likely to be in a high strain
rate regime (geophysical strain rates tend to be 10 orders of magnitude lower,  see Figure 2 by \citealt{hammond13}, 
\citealt{nimmo04b}  and discussion below).

The above estimate for the strain and strain rate used an impulse approximation, assuming that
the body does not have time to elastically respond during the encounter.   We compare the
elastic response time scale to the encounter time scale 
\begin{equation}
\frac{t_{enc}}{t_{elas}} =   \left( \frac{q}{R} \right)  \left( \frac{v_c}{V_q} \right) 
\left( \frac{E}{e_g} \right)^\frac{1}{2} \label{eqn:ratio}
 \end{equation}
If $t_{enc}/t_{elas} \ll 1$ then the impulse approximation is valid, whereas if $t_{enc}/t_{elas} \gg 1$
the encounter can be considered adiabatic.  When the ratio of time scales is near unity, the impulse approximation can be used
but the response should be reduced by a factor that depends on the ratio of the two time scales.

In Table \ref{tab:bodies} we list gravitational energy densities $e_g$,  time scales $t_{grav}$ and densities for
some moons  that exhibit long faults, chasmata or graben complexes.   The planet Mars, which exhibits 
the extremely prominent chasm, Valles Marineris, is included for later discussion.
The gravitational time scales in the icy bodies range from about 1000-2000~s (15 to 30  minutes).
We compare the gravitational binding energy densities, $e_g$, to an estimate for the Young's modulus of ice. 
Observations of ice shelf response to tides on Earth give an effective Young's modulus $E_{ice} \sim 0.9$~GPa \citep{vaughan95}, 
an order of magnitude below that of solid ice in the lab. 
On icy moons, porosity and surface fracturing may  lower the effective value of the Young's modulus 
(\citealt{nimmo06}; see \citealt{collins10}  for a review).  
The values of gravitational energy density
$e_g$ for icy bodies in Table \ref{tab:bodies} range from 0.3 GPa for Tethys to 2.14 for Titania.  Depending upon the value
 adopted for the Young's modulus for ice, the ratio $e_g/E_{ice}$ ranges from about 1  to 0.1 for these bodies. 
 
 Using the estimate for the ratio $e_g/E_{ice}$ we estimate the strain during an encounter.
For an equal mass and density perturber, the pericenter distance must be more than twice the body's radius; $q  > 2R$.
For a parabolic grazing encounter with $q=2R$, the pericenter velocity is equal to the escape velocity
at pericenter, and  $V_q =  \sqrt{2} v_c$ with $v_c$ defined in equation \ref{eqn:vc}.
Equation \ref{eqn:strain2} then implies that the strain caused by the tidal encounter could be of order 10\% for a parabolic  equal mass encounter.  
The ratio $E_{ice}/e_g$ is large enough that the time scale for the encounter could be a few times longer
than the elastic time scale.  This should reduce the strain on the surface compared to that estimated
using the impulse approximation in equation \ref{eqn:strain2} by a factor of a few 
(with factor that depends on the ratio ${t_{enc}}/{t_{elas}}$). 
Icy moons are likely to contain rocky higher density and strength cores and here  
 we have neglected compositional variation in our estimate of the tidally induced surface strain.
 This too would reduce the estimated strain value by a factor of a few.
Nevertheless our low estimated ratio of $E_{ice}/e_g$ implies that icy crusts or mantles in these icy bodies are weak enough
that deformation at the level of few percent strain is expected in close parabolic tidal encounters with similar mass objects. 

Close encounters with a similar mass body would be in a regime of high strain rate, compared
to most geophysical settings for ice, but lie in the mid to lower end of laboratory measurements \citep{lange83,schulson99,fortt12}. 
Figure 2 by \citet{petrovic03} illustrates that
the  tensile strength of ice is relatively insensitive to the strain rate, however \citet{lange83} found
a dependence, with the strength increasing at very high strain rates. 
At strain rates of order $10^{-6} {\rm s}^{-1}$
the tensile strength of ice 
is a few MPa \citep{lange83,schulson99,petrovic03}.
Using a Young's modulus of a few GPa,  brittle failure is likely to 
take place  if the strain under uniaxial tension exceeds 0.01 - 0.1\%. 
This crude strain based estimate for brittle failure ignores a dependence of the material strength on depth or pressure.
One of the modes of tensile failure 
causes fractures parallel to one of the axes of principal stress and is independent of confining pressure
\citep{jaeger76}.  This is equivalent to the limit with principal stress $\sigma_1 \sim 0$
(as on the body's surface), $\sigma_2 <0$, corresponding to extension,  and the Griffiths failure criterion that is a function 
only of the uniaxial tensile strength.  Here fractures are expected parallel  to the direction associated with $\sigma_3$ (on the surface
and perpendicular to the principal axis of tensile stress).
Based on our estimate of the induced strain during a close encounter with a massive body (equation \ref{eqn:strain2})
and using the values for gravitational energy density
$e_g$ for the icy bodies listed in Table \ref{tab:bodies}, we estimate that tidal stress and associated body
deformation 
could cause sufficient tension on the surfaces of these bodies to exceed the tensile strength of their icy crusts and 
cause widespread surface brittle failure.
 

\begin{table}
\vbox to 80mm{\vfil
\caption{\large  Bodies exhibiting long chasma or graben complexes \label{tab:bodies}}
\begin{tabular}{@{}lllllll}
\hline
                & mass                & radius & $e_g$ & $t_{grav}$ & $\rho$\\
                & ($10^{21}$ kg) & (km)   & (GPa)  & (s)            & g~cm$^{-3}$     \\
 \hline
 Tethys     &  0.62               &  531      &   0.32  & 1906 & 0.98 \\  
 Dione     &1.1                    &     561   & 0.81 &   1556  &1.48 \\ 
 Rhea      &  2.3                  &    764     &  1.04    & 1701 & 1.24 \\ 
 Ariel        &  1.3                  &     579    &  1.08     & 1468  & 1.66 \\ 
 Titania     &   3.5                &    788      &  2.14   &  1443 & 1.72 \\ 
  Charon   &  1.5                & 603      & 1.16   & 1472  & 1.65 \\ 
 \hline
 Mars        &  642                &  3389     & 208 & 953 & 3.93 \\
 \hline
\end{tabular}
{\\ The gravitational energy density $e_g$ is computed using equation \ref{eqn:eg} and 
the gravitational time scale $t_{grav}$ is computed using 
equation \ref{eqn:tgrav}.  Also listed are the mean densities.
Tethys, and Dione  are moons of Saturn, Ariel and Titania are moons of Uranus, and Charon is a moon of Pluto.
}}
\end{table}

\section{Mass-Spring Model Simulations}

Because of the short time scale of tidal encounters we model brittle-elastic behavior alone 
and neglect ductile, plastic or liquid-like behavior.  We model only the tidal encounter, and do
not model longer time scale viscoelastic behavior, such as crustal flexure, subsidence and decompression melting.
Mass-spring models or lattice spring models are a popular method for simulating soft  elastic bodies
\citep{meier05,nealen06}.
In  a mass-spring model, massive particles are connected with a network of massless springs.
Mass-spring models are considered a better choice than a finite difference
method when a fast, but not necessarily
accurate simulation is desired. However, with enough masses and springs 
and an appropriate choice of spring types, behaviors and geometry for the network of  connections, 
mass spring models can accurately represent elastic materials 
\citep{hrennikoff41,monette94,ostoja02,clavet05,kot15}.
Tidal spin down of spherical bodies can accurately be modeled with a random mass-spring model \citep{tide}.  

Our simulations differ from those used to model granular materials (e.g., \citealt{richardson09,richardson11,schwartz13}).
We apply gravitational, elastic and damping forces  to pairs of particles with
the direction of force  parallel to the vector connecting the pair.
In contrast  in simulated granular materials, particles that touch can exert a torque on each other.
In a mass-spring model ductile and plastic behavior is only possible if spring rest lengths can vary or if
springs can dissolve and reform (e.g., \citealt{clavet05}).   In contrast, simulated granular materials
can flow like a strengthless incompressible fluid (e.g., \citealt{leinhardt12}).

To study crustal deformation we need to numerically resolve the 
surface with a moderate number of surface particles. 
The spherical surface is supported by the interior of the body and 
feels stresses due to deformation of the interior.
We model the surface using a regular
2-dimensional triangular lattice of masses.  The lattice structure allows us to track and render the surface. 
We model the interior with randomly distributed more widely spaced and more massive particles than used
in the surface shell.    

Most mass-spring models do not take into account the self-gravity of the body itself. 
Our mass-spring model is embedded inside an N-body code and so takes into account gravitational forces
between all particles in the body itself as well as the perturbing body.
To carry out our simulations we use the  modular code \texttt{rebound},  
an open-source multi-purpose N-body code for collisional dynamics \citep{rebound}
(available at \url{http://github.com/hannorein/}).
From the \texttt{rebound} version 1 code (from June 2015), 
we use the open-GL display, open boundary conditions, the direct all pairs gravitational force computation and the 
leap-frog integrator (second order and symplectic) to advance particle positions.  
To the particle accelerations we have added additional spring 
and hinge forces, as described below.
We have modified the display to illustrate the springs between particles and add textures to the surface layer.

We work in units of the planetesimal body radius and mass $R=1, M=1$.
Time is specified in units of $t_{grav} = \sqrt{R^3/GM}$ (equation \ref{eqn:tgrav}) and we refer to this
as a gravitational time scale.
In these units, the velocity of a particle in a circular orbit just grazing the surface of the body
is $1$, and the period of this orbit is $2 \pi$.  
Velocities are given in units of $\sqrt{GM/R}$, accelerations in units of $GM/R^2$ and spring constants
in units of $GM^2/R^3$.
Pressure, energy density and elastic moduli are given in units of $GM^2/R^4$ or $e_g$ (equation \ref{eqn:eg}).
In units of $R=1, M=1$ the mean density of the primary body is $\bar \rho = 3/(4\pi) = 0.239$.

The parameters used to describe  our simulations are summarized in Table \ref{tab:sim}.

\subsection{Spring forces}

The code \texttt{rebound} advances particle positions using their accelerations \citep{rebound}.
We add spring and hinge forces as
additional forces by adding to the particle accelerations
at each time step.
To compute the particle accelerations, 
the forces from each spring and hinge on each particle are divided by the mass of each particle.

The elastic force from a spring between two particles $i,j$ with masses $m_i, m_j$ and 
coordinate positions ${\bf x}_i$, ${\bf x}_j$,  
on particle $i$ is computed as follows.
 The vector between the two particles
${\bf x}_i - {\bf x}_j$ gives a spring length $L_{ij} = |{\bf x}_i - {\bf x}_j| $ that we compare with 
the spring rest length $L_{ij,0}$.
The elastic force from a spring between two particles $i,j$ on particle $i$ is computed as
\begin{equation}
{\bf F}_{i}^{elastic} = -k_{ij} (L_{ij} - L_{ij,0}) \hat {\bf n}_{ij}  \label{eqn:Fe}  
\end{equation}
where  $k_{ij}$ is the spring constant and the unit vector $\hat {\bf n}_{ij} = ( {\bf x}_i - {\bf x}_j )/ L_{ij}$.  
The force has the opposite sign on particle $j$.
When the spring length is longer than its rest length $L_{ij}$ the force pulls the two masses together
and when it is shorter than its rest length the force pushes the two masses apart.

The strain rate of the spring is
\begin{equation}
 \dot \epsilon_{ij} = \frac{\dot L_{ij} }{  L_{ij,0} }= \frac{1}{L_{ij} L_{ij,0}} ({\bf x}_i - {\bf x_j}) \cdot ({\bf v}_i -  {\bf v}_j) 
 \end{equation}
where ${\bf v}_i$ and ${\bf v}_j$ are the particle velocities,  $\dot L_{ij} $ is the rate of change of the spring length, 
$L_{ij}$ and $\epsilon_{ij} = (L_{ij} - L_{ij,0})/L_{ij,0}$ is the spring strain.
To the elastic force on particle $i$ we add a damping force proportional to the strain rate
\begin{equation}
 {\bf F}_{i}^{damping} = - \gamma_{ij}  \dot \epsilon_{ij} L_{ij,0} \mu_{ij} \hat {\bf n}_{ij}  \label{eqn:dampforce}
 \end{equation}
with damping coefficient $\gamma_{ij}$  that is equivalent to the inverse of a damping or relaxation time scale.
The coefficient  
$\gamma_{ij}$  is independent of the spring constant $k_{ij}$.
Here $\mu_{ij}$ is the reduced mass $\mu_{ij} \equiv m_i m_j/(m_i + m_j)$. 
The damping force is in parallel with the elastic term so the spring model 
approximates a viscously damped elastic material
or a  Kelvin-Voigt solid (see \citealt{tide}).  Between each pair of particles, forces are oriented along the vector spanning the two particles
and this ensures angular momentum conservation.

To maintain numerical stability, the time step should be smaller than the time it takes physical information to travel
between adjacent mass nodes.    The ratio of the P-wave velocity to gravitational velocity $V_{P}/v_c \sim \sqrt{E/e_g} $
and this exceeds 1 otherwise  the body would collapse due to self-gravity.
In our simulations the speed of elastic waves exceeds $v_c$ so we must choose the time step to be shorter
than  the time it takes for elastic waves to propagate between adjacent mass nodes;
\begin{equation} 
dt \la {\rm min}_{i,j} \sqrt{ \frac{\mu_{ij}}{k_{ij} b}  } \label{eqn:dt}
 \end{equation}
 taking the minimum value of all springs in the model and with factor $b$ the number of springs per node.

\subsection{Random isotropic mass-spring model for the interior}
\label{sec:mush}
 
We generate an isotropic random mass-spring model for the interior similar to the random
spring model described by \citet{kot15}.   
Each mass is a node and all nodes in the interior have the same mass.
Random particle positions are generated but a  particle is added to the network only if it is sufficiently
separated from other nodes (at a distance greater than minimum distance $d_{IS}$).  
Springs are added between two nodes if the distance between nodes
is less than distance $d_I$.   Such a spring network is isotropic, has Poisson ratio $\nu=1/4$ 
and has a Young's modulus of
\begin{equation}
 E_I 
\approx \frac{1}{6V} \sum_i k_i L_i^2  \label{eqn:Emush}
\end{equation}
\citep{kot15},
where $k_i$ and $L_i$ are the spring constants and spring rest lengths for spring $i$, and $V$
is the total volume.  In the above expression the sum is over all springs in the interior volume.

We compute the velocity of $P$ waves in the interior with
\begin{equation}
V_{I,P} = \sqrt{E_I/\rho} \left(\frac{1-\nu}{(1+ \nu)(1-2\nu)} \right)^{1/2} \approx 1.1 \sqrt{E_I/\rho} \label{eqn:V_IP}
\end{equation}
With non-zero damping coefficients, the stress is the sum of an elastic term proportional to the strain and
a viscous term  proportional to the strain rate.   The bulk and shear viscosities  can be estimated from the damping
coefficient, $\gamma$,
and other integrated properties of the mass spring model \citep{tide}.

In a three-dimensional volume, the number of springs is proportional to the number of particles in the interior, $N_I$.
The mean length of the springs $L \propto N_I^{-1/3}$ giving spring constant $k \propto E_I L \propto N_I^{-1/3}$ to maintain a specific 
Young's modulus.   The mass of each particle is $m \propto 1/N_I$.  Altogether this gives a time step (using
equation \ref{eqn:dt})
$dt \propto N_I^{-1/3}$ and, as expected, the more particles simulated, the smaller is the required time step.
\citep{tide} has tested sensitivity of the random spring model to numbers of simulated mass nodes and numbers of springs
per node. 

\subsection{Initial conditions for the interior: Stretching the springs and strengthening the core}
\label{sec:stretch}

We first generate springs with
 rest length equal to the distance between the spring vertices.
However this does not generate an equilibrium state for our body because of compression due to self-gravity.
When the simulation is begun in this state the body shrinks and then bounces, eventually damping
to  a  denser equilibrium state than the initial condition.  To begin with the system nearer equilibrium we start
with the springs initially slightly under compression so that they counter-act self-gravity.
We iteratively stretch all springs by the same amount to zero the acceleration at the surface.
We then increase the spring strengths in the center of the body so as to 
approximate a state of hydrostatic equilibrium.
A constant density self-gravitating sphere in hydrostatic equilibrium has pressure as a function of radius
\begin{equation}
P(r) = \frac{2}{3} \rho^2 G \pi (R^2-r^2) \end{equation}
(consistent with equation 2 by \citealt{dobro90}).
The pressure at any radius is approximately $P \sim E_I \epsilon$ where $\epsilon$ is the strain
of each spring. 
The resulting model is nearly in equilibrium, nearly constant density, and does not bounce excessively at
the beginning of the simulation.  

\begin{figure*}
\includegraphics[width=7.1in]{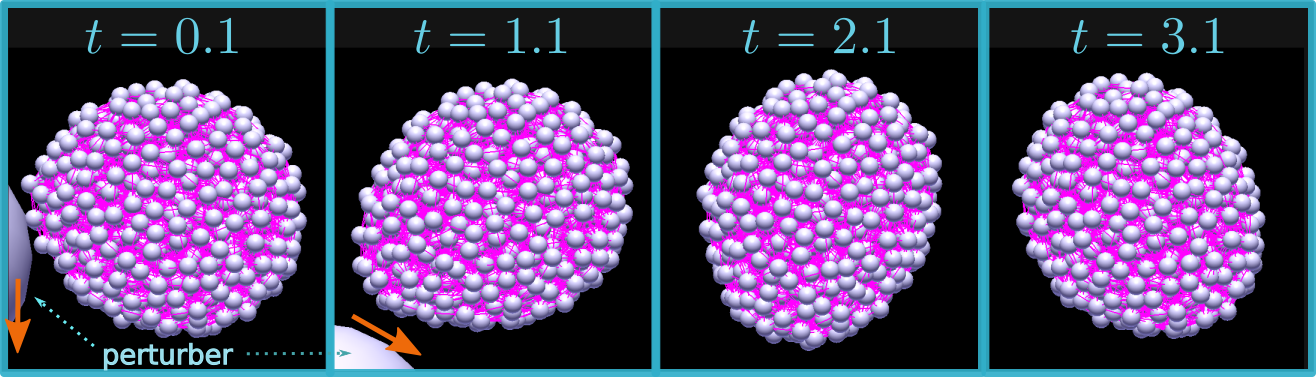} 
\caption{Simulation of a near parabolic tidal encounter of a random spring elastic model with parameters
listed in Table \ref{tab:sim_list} and \ref{tab:sim_qlist} for the N simulation.
 The perturbing body approaches from the top left,
 and is seen in the leftmost two panels.
Different times in the simulation are shown from left to right, separated in time 
by 1 in units of $t_{grav}$ (equation \ref{eqn:tgrav}) and labelled by time from pericenter.  
The leftmost panel shows a time just after closest approach.
The primary body is simulated with a mass-spring model and springs are shown as pink connecting
lines between particles.  The body is initially spherical but is elongated by the tidal force of the perturber.  
 Vibrational oscillations are excited in the body by the tidal encounter.
 The rendered spheres are shown to illustrate the random distribution of node point masses,
 not imply that the body behaves as a granular rubble pile (e.g.,  \citealt{richardson09}).  
}
\label{fig:mush}  
\end{figure*}

To illustrate the elasticity of the interior we show a gravitational tidal encounter
with a perturber mass $M_2= M$, equal to the primary body, $M$.  The simulation
is shown in Figure \ref{fig:mush} and has body parameters listed in Table \ref{tab:sim_list} 
and encounter parameters listed in Table \ref{tab:sim_qlist} under the row N. 
The perturbing mass, $M_2$, is modeled as a solid sphere
that does not deform during the encounter and it is rendered as a sphere with the same density as the primary body.
The simulation view has been shifted so that it remains in the center of mass frame of the primary body.
As the spring damping coefficient is low, the body is soft, like jello and
 the body continues to vibrate and deform  after the tidal
encounter.   Vibrational oscillations are excited in the body by the tidal encounter.
We have checked that the elasticity code conserves momentum and angular momentum, 
as would be expected as particle interaction forces (including damping forces)
are  radial.\footnote{The difference between angular momentum at the beginning and end of the simulation is
less than $10^{-12}$ in our N-body units described in section 2.0.}  With the resolved body in a circular orbit about
 a larger mass and on long time scales, the spring damping causes the resolved
 body to spin up or spin down, as expected, and with rate matching that predicted analytically  \citep{tide}.

\subsection{Crustal Shell Model}
\label{sec:shell}

\begin{figure}
\includegraphics[width=3.3in]{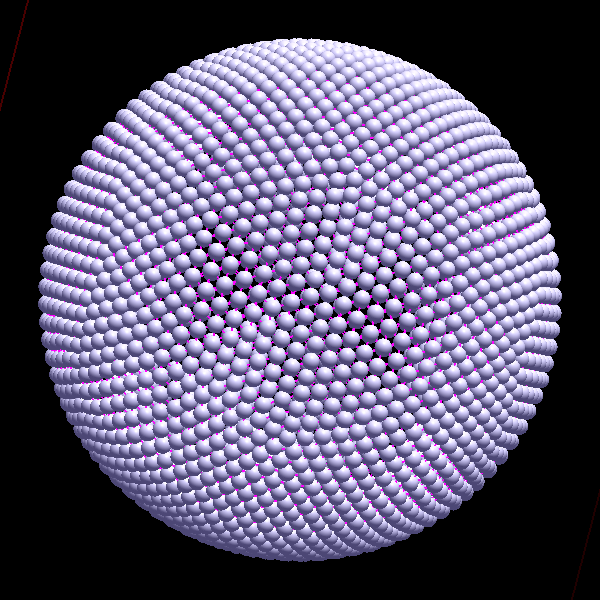}
\caption{Initial shell model that is based on a subdivided icosahedron.
A triangular spring lattice connects the vertices.
}
\label{fig:shell} 
\end{figure}

We model the elastic crust with a  2-dimensional spherical lattice. 
Each vertex of the shell lattice is a mass node and each node has the same mass.  However the shell node masses
are not the same as the interior node masses.
To create a sphere of particles we begin with vertices of an icosahedron.
Each face is then subdivided into 4 triangles. 
We recursively repeat this operation until we reach a desired total number of vertices in the shell, $N_S$.
Each vertex is the same distance from the body center.
After subdivision, springs are placed  with spring constant $k_S$
between each adjacent vertex, creating a triangular
spring lattice (see Figure \ref{fig:shell}) and giving the 2-dimensional shell elasticity (e.g. \citealt{monette94,vangelder98}).
The vertices of each triangular face are stored to later aid in displaying (rendering) the body's surface.  

For a triangular 2D lattice, the Young's modulus depends on the spring constant 
\begin{equation} E_{2D} = \frac{2}{\sqrt{3}} k_S \end{equation}
and the Poisson ratio is $\nu_{2D} = 1/3$  (see  equations 3.7 by \citealt{monette94} 
or equations 10,12 by \citealt{kot15}).
The modulus is a force per unit length rather than a force per unit area as is true in 3D.
The sphere locally contains a single layer of particles and approximates a thin plate with
\begin{equation}
E_{2D} = E_S h_S \label{eqn:E23}
\end{equation}
where $E_S$ is the Young's modulus of the plate and $h_S$ its thickness.
For $N_S$ particles in the shell, the  mean length of the springs $L_s \propto N_S^{-1/2}$, the mass of
each particle $m_s \propto N_S^{-1}$ but $k_S$ is independent
of the number of particles.   Hence the required time step (using equation \ref{eqn:dt}) 
$dt \propto N_S^{-1/2}$. 
The speed of P-waves in the shell is 
\begin{equation}
V_{S,P} \approx \sqrt{E_{2D}/\Sigma_S}  \sim \sqrt{\frac{k_S 4 \pi R^2}{M_S} }\label{eqn:V_SP}
\end{equation}
where $\Sigma_S$ is the mass per unit area in the shell.

\subsection{Hinge Forces and Flexure of the Crustal Shell}
\label{sec:hinge}

As a 2-dimensional lattice is used to simulate the body's crust, 
the simulated crust is  a thin membrane and when simulated with radial elastic forces between particles,
 and in the presence of gravity, it would be unstable and would bend and sag.  
Using the triangular phases in our crustal sphere, we use a hinge model to resist bending of adjacent triangular
faces.  

\begin{figure}
\hspace*{1cm}
\includegraphics[width=3.0in, trim=1cm 0 0 0]{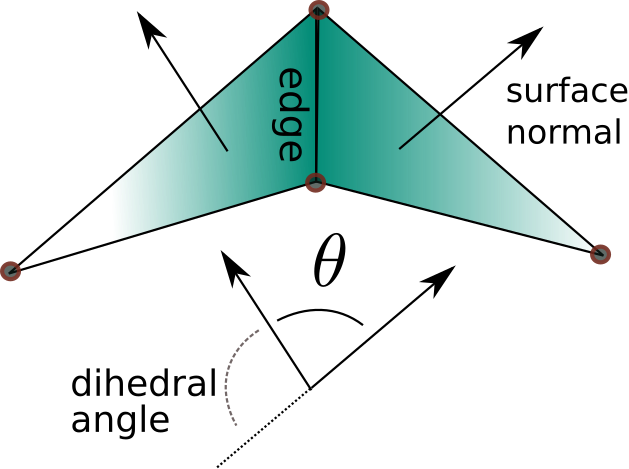}
\caption{Hinge model for crustal flexural stiffness as described by \citet{bridson03,grindspun03}.
For each edge in the crustal shell, the dihedral angle is used to compute 
the bending force on each vertex particle.
}
\label{fig:hinge}  
\end{figure}

The force on each edge depends on the dihedral angle between the two faces 
as described by \citet{bridson03} in their section 4, see Figure \ref{fig:hinge}, also see \citet{grindspun03}. 
An angular dependent force is applied to the two vertices in a spring edge 
and to two adjacent vertices forming the hinge.
The magnitude of the applied force is
\begin{equation}
 |F| =  k_e \frac{L^2}{|N_1| + |N_2|} \sin ((\theta - \theta_{rest})/2)  \label{eqn:F_hinge}
 \end{equation}
where $L$ is the length of the edge.   The dihedral angle is equivalent to $\pi - \theta$
with angle $\theta$ between the two vectors, ${\bf N}_1$ and ${\bf N}_2$, that
are normal to the surfaces of each triangular face.
Here $|{\bf N}_1|$ and $|{\bf N}_2|$ are the areas
of the two triangular faces  of the hinge (see Figure \ref{fig:hinge}).
The force strengths are the same on each vertex but applied with direction
given in vector form given by \citet{bridson03} assuring momentum and angular momentum conservation.
The angle $\theta_{rest}$ is a rest bend angle and describes the angle of the surface without stress. 
We adjust the sign of $\theta$ by ordering the vertices in the edge. 
Before the simulation starts for each hinge we compute $\theta_{rest}$ from the initial configuration of the surface lattice. 
Here our surface lattice is spherical but rest hinge angles can be computed for any smooth 2-dimensional lattice. 
This ensures that the initial state of our simulated crustal shell is not under flexural stress.
As is true for the spring forces, accelerations for each mass are computed from the applied forces
by dividing by the mass of each vertex particle.

The coefficient $k_e$ is in units of force but the force is applied at each hinge and with opposite sign in the center
and ends of the hinge.   The hinge is similar to a flexed beam, 
held at both ends but with an applied force at its center, and with
width equal to the length of the hinge edge and length equal to the distance between outer vertices.
The hinge model locally approximates a thin plate with Young's modulus $E_S$ and plate thickness $h_S$ with 
\begin{equation}
\frac{ k_e L_S}{2} \sim \frac{E_S h_S^3}{12(1 - \nu^2)}  \equiv  D_F  \label{eqn:D_F}
\end{equation}
with $L_S$ the length of an edge and $D_F$ equal to the  flexural rigidity or bending stiffness of the plate 
(e.g., \citealt{batty12}).  This is approximate as simulated static load tests have not been carried out
with the membrane/hinge model  (though simulated static beam tests have been done
for the lattice and random spring models, see \citealt{kot15}).
  


 
\subsection{Crack formation}
\label{sec:crack}
 
Crack formation in the crustal shell can be simulated in a spring model by breaking springs that have a strain value
 above a critical threshold value, $\epsilon_S$, (e.g., \citealt{norton91,marder93,hirota00,sad11}).  
The springs in our simulated spherical shell behave as perfect linear springs
 until the moment that they reach the threshold value.
If we completely dissolve failed springs, eventually three particles in a hinge triangle face (see Figure \ref{fig:hinge})
 could approach a line.  When this happens the hinge  force (equation \ref{eqn:F_hinge}) becomes infinite and
 the surface is unstable.  Rather than completely dissolve springs,
 we instead reduce the spring force constant  by a factor $F_{Sk}$.
 
When a crack forms in a solid, the stress perpendicular to the crack's path is reduced. 
 Subsequent deformation concentrates stress at the new crack tip, which in turn fails, allowing the crack to propagate. 
 In a simulation, the stress is redistributed as vertex positions are updated, and this allows the crack
 to propagate.    A simple way to allow the stress redistribution (or relaxation) to take place within
 the simulation is to allow only 
 only one element to rupture in a given time interval, often the simulation time step (see discussion by \citealt{pfaff14}).
Here, we only allow a single shell spring, that with maximum strain, to fail in a given time interval, denoted $t_{fail}$, which 
we set between one and three time steps. 
This relaxation procedure allows linear fractures to propagate, and prevents large areas of the surface
 from failing simultaneously.  
A more sophisticated code would  allow more than one crack to simultaneously propagate 
(and in this case additional relaxation steps must be implemented and the residual momentum propagation 
taken into account in each 
crack tip region; \citealt{busaryev13,pfaff14}).

\begin{figure}
\includegraphics[width=3.5in]{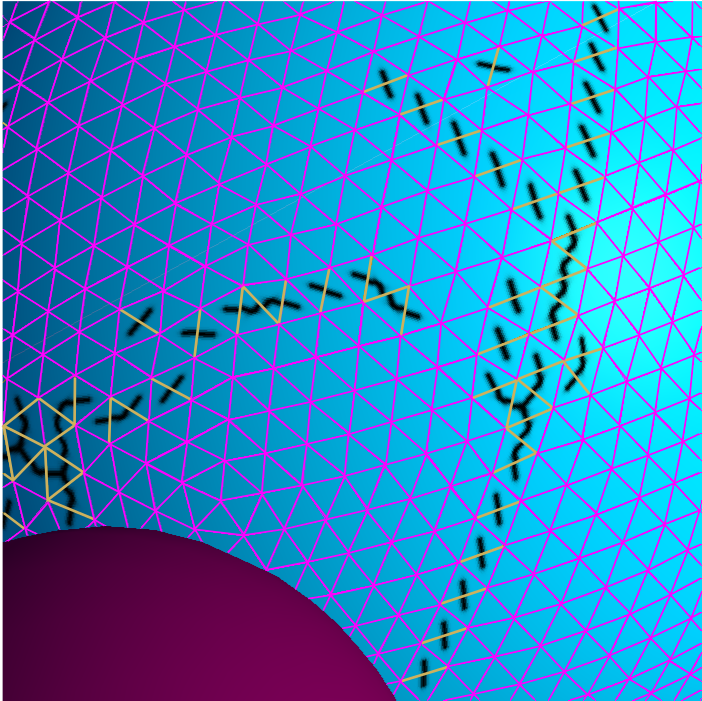} 
\caption{Rendering of crustal failure.  The crustal shell is modeled with a triangular lattice with
springs connecting each lattice point.   Lattice points are mass nodes in the spring network. 
Springs (along the edges of triangles in the mesh) that have exceeded their maximum strain are colored tan instead of pink.  
Each surface triangle is displayed based on the properties of 
its edge springs.  If an edge spring has failed, a black bar is shown on the triangular face perpendicular to the failed spring.
If two edge springs fail, then a bent black bar is shown on the triangle with each end touching  a failed
spring.  If all three
edge springs have failed then a three pronged black fork is shown on the triangular face.
If all springs in the surface network fail, the lattice that is dual to the triangular lattice, a hexagonal one, is seen in black.
Long connected or partially connected sets of black bars illustrate fractures in the surface shell. 
In this close-up view tidal forces from the perturber
(colored magenta) has caused some springs in the surface lattice to fail.
}
\label{fig:render}  
\end{figure}

 
The surface of the shell is displayed using the triangular faces from each lattice triangle. 
Each triangular face is displayed
 with a texture.   The edges of each triangle contain connecting springs.  The texture displayed
on each triangular face depends on the number and orientations
 of failed edge springs (see Figure \ref{fig:render}).   Long connected or partially connected sets
of black bars illustrate surface fractures.   The surface lattice introduces directional biases in the crack rendering.
Offset individual black bars in Figure \ref{fig:render} are artifacts from the lattice rather than en-echelon structures.  

\subsection{Support of the Shell}
\label{sec:cross}

The crustal shell lattice must be supported by the interior otherwise it will collapse due to gravity.  
To connect the two components of our 
models (amorphous interior to the lattice shell membrane) we insert springs between shell particles and interior particles.
We create springs for all mixed pairs of particles, a pair consisting
of  a particle in shell and a particle in the interior, that have inter-particle distance less than 
a distance $d_C$.
The springs that cross between the interior and shell, we denote {\it cross} springs and they are
described with a spring constant $k_C$, number $\NS_C$ and mean rest length $L_C$.  
The number of cross springs per shell particle can be increased by increasing the distance $d_C$.

\begin{figure}
\includegraphics[width=3.5in]{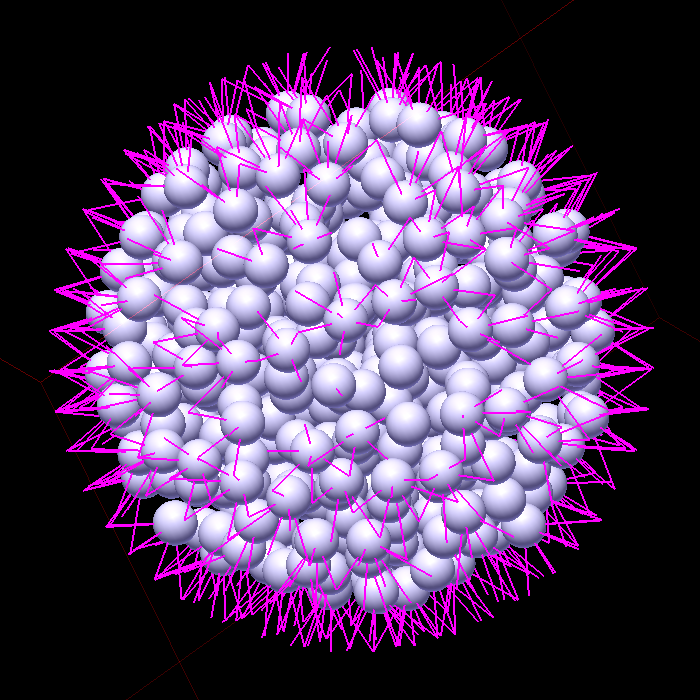} 
\caption{Cross springs (shown in pink) connect the shell and interior particles.
Only the interior particles are shown as spheres.
}
\label{fig:cross}  
\end{figure}

To ensure that the initial model does not bounce excessively at the
beginning of the simulation, these springs are initially set slightly under compression, as  described
in section \ref{sec:stretch}.  An illustration of our cross springs is shown in Figure \ref{fig:cross}.

We match the speed of elastic waves traveling horizontally in the shell to the speed of vertical 
oscillations of the cross springs.
The strength of the cross springs we estimate using the dimensional scaling of equation \ref{eqn:Emush}
\begin{equation}
k_C \sim \frac{E_I}{L_C} \frac{N_S}{\NS_C}  \end{equation}
where the Young's modulus is that of the interior and the value for spring constant $k_C$ depends on the number
of cross springs per shell particle.
Compression is communicated elastically from the crust to the interior with a  speed
\begin{equation} 
V_{Co} = \sqrt{\frac{k_C \NS_C}{M_S}} L_C 
\label{eqn:V_C}
\end{equation}
where $M_S$ is the mass in the shell.  
We adjust $d_C$ and $k_C$ so that the coupling speed $V_{Co}$ is comparable to the velocity of P-waves
in the interior, $V_{I,P}$.

\subsection{Damping to an equilibrium state}
 
If the crustal shell is initially under compression at the beginning of the simulation, 
increased extension would be needed to cause spring failure.
If there are local
crustal stresses present at the beginning of the simulation, they would influence the location of tensile failure during the tidal encounter.
Despite adjustment of interior spring lengths and strengths (described in section \ref{sec:stretch}), and setting
the cross springs slightly under compression, 
the  body bounces radially at the beginning of a simulation.  
This can affect the timing (and so location) of crustal tensile failure, as failure would be most likely to occur 
when the body has largest radius.
 We would like shell springs to be  at their rest lengths at the beginning of the simulation and the body should
 be in equilibrium and not vibrating.  To ensure equilibrium and zero the stress in the crust we
 run a relaxation simulation before each tidal encounter.
 The relaxation simulations are run 
without a tidal perturber, of the primary body alone, with a large
damping parameter $\gamma$. 
During this relaxation simulation,  springs are not allowed to fail and
 we slowly adjust the lengths of the shell springs so that they approach a zero force condition.
The rest hinge angles are periodically reset to zero the flexural stress of the surface.

At the end of the relaxation, shell springs are very close to their rest lengths (under no extension or compression), the vibrations
of the body have decayed, and the crustal membrane is under no flexural stress.  The support of the crustal
membrane by compression by the cross springs approximates hydrostatic support even though there is no density contrast.
The relaxation simulations are run $T = 3$ (in units of the gravitational time scale; equation \ref{eqn:tgrav})
under a damping coefficient (for all springs) of $\gamma = 50$.
After the relaxation run is finished we store the positions and velocities of all masses and properties of all springs and hinges
so that they can be read back into the code to run the tidal encounters.
By simulating the relaxed body with a low damping coefficient and without an encounter, we 
  check that the body is stable and that springs do not fail in the absence of any perturber.

\subsection{Parameter choices for the Tidal encounters}
\label{sec:choice}

To describe the primary body a large number of parameters  must be chosen (see Table \ref{tab:sim}).  
We adjusted the minimum distance distance
between particles $d_{I}$ to be somewhat larger than the distance between shell particles  as we need to well resolve
the shell but not the interior. 
The parameter $d_{IS}$ was adjusted so that the number of springs per node in the interior exceeded 10, and
so that the effective elastic coefficients (Young's modulus and Poisson ratio)
are not strongly dependent on the spring network \citep{kot15}.
 The shell mass and distance between shell and outer boundary of the interior 
 were adjusted to give the shell approximately the same density as the interior
and so that implied shell thickness $h_S$ is approximately
consistent with that estimated from the flexural strength and speed of elastic waves through the crust.

The spring constant in the interior was chosen so that
the Young's modulus of the interior is approximately 3, matching an estimate for the value of ice's Young's modulus
in units of $e_g$ for the icy bodies listed in Table \ref{tab:bodies}.    
We adjusted spring constants
in the cross springs and shell so that the speed of elastic waves in the shell is similar
to that in the interior and the speed of vibrations passing from shell to cross spring $V_{S,P} \sim V_{Co} \sim V_{I,P}$
and assumed composition of crust and interior is similar. 
The spring constant in the shell was adjusted so that the implied crustal thickness from equation \ref{eqn:E23}
gave $h_S \sim 0.02$ (in units of radius), corresponding to 10 km crustal thickness for a body with of radius $R= 500$ km.
Estimates for the thickness of the icy crust  on Dione and Tethys range from 1 - 7 km, and
are based on assuming that topographic features are signatures  of flexure of a broken elastic plate 
\citep{giese07,hammond13}.
Our modeled thickness exceeds these estimates by a factor of 2--3 but they are based on flexure over long time scales (Myr)
and  the effective elastic thickness of the crust should be larger on the shorter tidal time scale (hours).
A comparison of observed and predicted flexural rigidity (proportional to effective thickness to the third power)
as a function of age of geological load for seamounts and oceanic islands implies that elastic thickness is a strong
function of strain rate (see section 6.7 by \citealt{watts01}) but it would be inaccurate to extrapolate 
over orders of magnitude to tidal time scales.
Our simulated crustal shell is connected via cross springs to the interior so vibrational waves can propagate
throughout the body, similar to the way seismic waves propagate through the Earth's mantle. 
Using the implied crustal thickness, $h_S$,
a flexural force parameter, $k_e$, was chosen to be  stronger than that estimated using equation \ref{eqn:D_F}) 
so as to maintain numerical stability (keep the hinges from collapsing) during the simulation.


 The strain value for surface spring failure for most of the simulations was chosen to be $\epsilon_S = 0.003$, 
 however weaker tidal encounters
would allow the surface to fracture were we to reduce this number.  
The strength reduction parameter $F_{Sk}$ (setting spring strength after failure in a shell spring) 
was set high enough to ensure that surface triangles never collapsed 
to a line during the simulation.  Collapse of a surface triangle causes the surface to become numerically 
unstable as the hinge forces become infinite
when the triangular face areas drop to zero.

After the relaxation runs are done, we run the tidal encounters. During the encounter simulations no new springs are created
and only shell springs are allowed to fail.  The tidal encounter simulations are begun and ended with the secondary mass 
located at distance of approximately 3
times the radius of the primary away from the  primary body center and are  run for a total time of $T \sim 3$ in
gravitational units.    The perturber is modeled as a point mass.
The inverse of our
gravitational time unit is equivalent to the spin of a body with surface near the centrifugal rotational breakup velocity.  
Most moons rotate much more slowly than this value (e.g., \citealt{M+D}).
As tidal encounters are fast (taking place on a time $t_{grav}$),  we ignore the role of the spin 
of the primary body, setting it to zero.
Gravitational softening
is set to 1/100 of the minimum initial inter-particle spacing and we have checked that its exact value
does not influence the simulations. 
Simulations were run at half the timestep listed in Table \ref{tab:sim_list} to check that
the resulting surface morphologies were similar and not  dependent on the time step.

\begin{table}
\vbox to170mm{\vfil
\caption{\large  Simulation Parameter Descriptions \label{tab:sim}}
\begin{tabular}{@{}lllllll}
\hline
$N_I$ & Number of particles in interior \\
$\NS_I$ & Number of interconnecting springs in interior \\
$L_I$  & Mean rest spring length in interior \\
$k_I$  & Mean spring constant of interior springs \\
$E_I$  & Young's modulus of interior \\
$d_I$    & Interior spring formation distance \\
$d_{IS}$ & Minimum initial interparticle distance \\
$V_{I,P}$  & Speed of P-waves in the interior \\ 
\hline
$N_S$ & Number of particles in shell \\
$\NS_S$ & Number of interconnecting springs in shell \\
$M_S$  & Mass of shell \\
$L_S$ & Mean rest spring length of  shell springs \\
$k_S$ &  Spring constant of shell springs \\
$k_e$  &  Bending spring constant for flexural strength of shell \\
$h_S$     & Simulated crustal thickness \\
$\epsilon_S$ & Maximum strain for spring failure in shell \\
$F_{Sk}$  & Factor spring constant $k_S$ is reduced after spring failure \\
$V_{S,P}$  & Speed of P-waves in the shell \\ 
\hline
$\NS_C$ & Number of cross springs connecting shell and interior  \\
$L_C$ & Mean rest spring length of cross springs \\
$k_C$  & Spring constant of cross springs \\
$d_C$ & Cross spring formation distance \\ 
$V_{Co}$ & Effective velocity of shell and interior coupling  \\
\hline
$\gamma$ & Spring damping coefficient \\
$dt$  & Time step \\
$t_{fail}$ & Time interval between failure of individual springs  \\
\hline
$M_2$ & Mass of tidal perturber  \\
$q$  & Distance between body centers at closest approach \\
$V_q$  & Relative velocity at closest approach\\
\hline
\end{tabular}
{\\ The first group of parameters describes the random elastic interior (Section \ref{sec:mush}).
The second group describes the shell lattice (section \ref{sec:shell} and \ref{sec:hinge}.  The third group describes
the cross springs connecting shell to interior (section \ref{sec:cross}).
The fourth group lists damping coefficient for all springs and the time step used in the integrations.
The last group lists properties of the tidal encounter.
The Young's modulus
$E_I$ is computed using equation \ref{eqn:Emush}. 
Lengths are in units of the body's initial radius.   Timescales
are in units of the gravitational time scale (equation \ref{eqn:tgrav}).
Masses are in units of the body's mass.
Moduli are in units of $e_g$ (the body's gravitational binding energy, equation \ref{eqn:eg}).
The spring damping coefficient is in units of inverse time.
Effective shell plate thickness is estimated using equation \ref{eqn:D_F} or \ref{eqn:E23}.
Velocities are computed using equations \ref{eqn:V_IP}, \ref{eqn:V_SP}, \ref{eqn:V_C}.
}}
\end{table}

\begin{table}
\vbox to130mm{\vfil
\caption{\large  Parameters for Simulations \label{tab:sim_list}}
\begin{tabular}{@{}lllllll}
\hline
            & N  &  T-series  \\  
           \hline
$N_I$    & 798   & 1869   \\
$\NS_I$ & 11632 & 23363  \\
$L_I$    & 0.30 & 0.176\\
$k_I$   &  0.0076 &0.076  \\
$E_I$   & 0.4  &   3.73   \\
$d_I$    &  0.38  &  0.10  \\
$d_{IS}$ & 0.15 & 0.23  \\
$V_{I,P}$   & 1.4 & 4.5   \\
\hline
$N_S$  &0 &2562 \\
$\NS_S$ &  &7680  \\
$M_S$   & &0.07   \\
$L_S$   & &0.075  \\
$k_S$   &  &0.06  \\
$k_e$   &  & 0.0075   \\
$h$       &   & 0.02 \\
$\epsilon_S$& & 0.003   \\
$F_{Sk}$  & &0.005   \\
$V_{S,P}$   & & 3.28 \\
\hline
$\NS_C$ &0& 5966  \\
$L_C$  & &0.156   \\
$k_C$  & &0.005   \\
$d_C$  & &0.18  \\ 
$V_{Co} $  & & 3.3  \\
\hline
$\gamma $ & 0.1 & 0.01  \\
$dt$  &0.003 & 0.002 \\
$t_{fail}$ & - & 3 $dt$ \\
\hline
%
\hline
\end{tabular}
{\\ %
Interior particle positions were 
randomly generated individually for each simulation and have slightly different
numbers of interior particles, interior and cross springs and $E_I$ from those listed here.
The T1 simulation has a slightly lower $\epsilon_S = 0.002$.
}}
\end{table}

\begin{table*}
\vbox to50mm{\vfil
 { \caption{\large Encounter Parameters for Simulations \label{tab:sim_qlist}}}
\begin{tabular}{@{}lllllll}
\hline
Simulation & $M_2$  & $q$  & $V_q$  & Description\\
\hline
N              & 1.0        & 2.13  & 1.84  & interior only\\
T5            & 0.5         &1.84 & 1.60 & fiducial \\
T1           & 0.1        & 1.47 & 1.62 & lower mass perturber and weaker crust \\
T10        & 1.0         & 2.10 &  1.61 & higher mass perturber  \\
T10b        & 1.0         & 2.22 &  1.94 & higher mass perturber and faster encounter \\
\hline
\end{tabular}
{\\ %
}}
\end{table*}

\begin{figure*}
\includegraphics[width=7in]{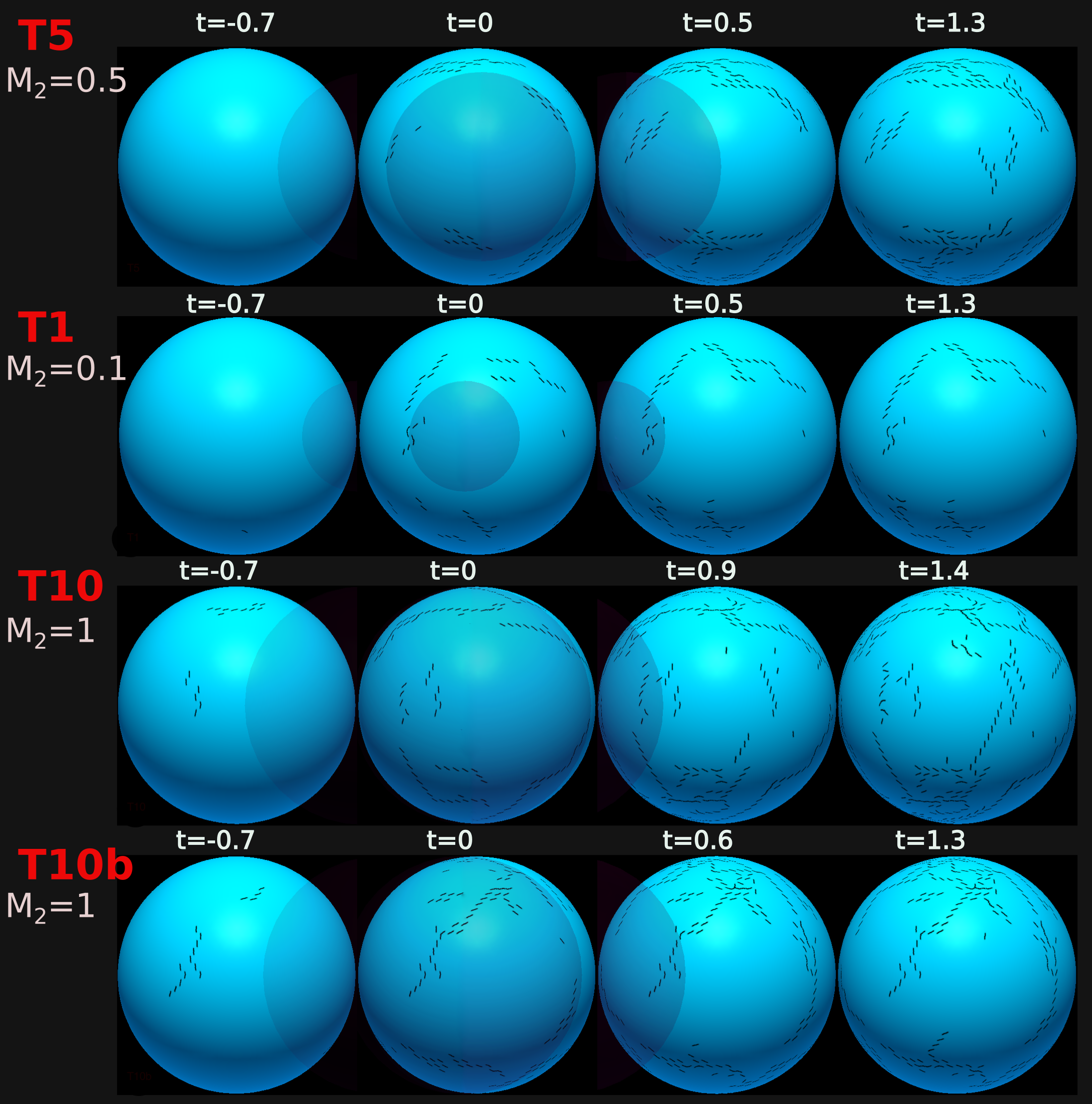} 
\caption{Simulation of tidal encounters  with parameters
listed in Table \ref{tab:sim_list} and encounter parameters listed in Table \ref{tab:sim_qlist}.  
From top to bottom the simulations are T5, T1, T10, and T10b.  Perturber masses are shown on the left.
Each sub-panel shows a different time with time advancing from left to right with times labelled from pericenter.  
Pericenter is approximately at the second column from left.   The line of sight is perpendicular to the encounter orbit plane.
As the body deforms in response to the tidal perturbation, long linear fractures appear on the surface.
}
\label{fig:T5}  
\end{figure*}

\begin{figure*}
\includegraphics[width=3in]{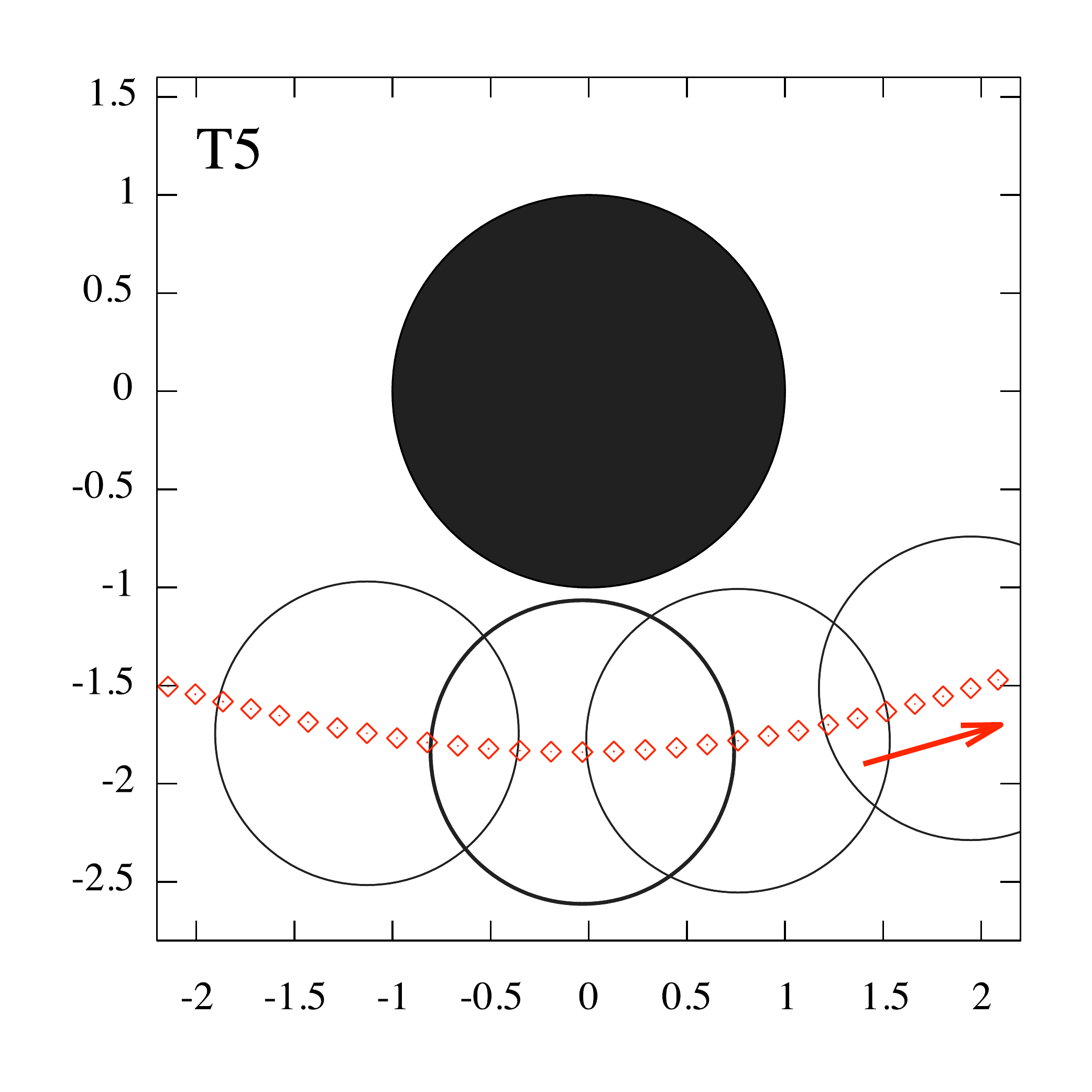} 
\includegraphics[width=3in]{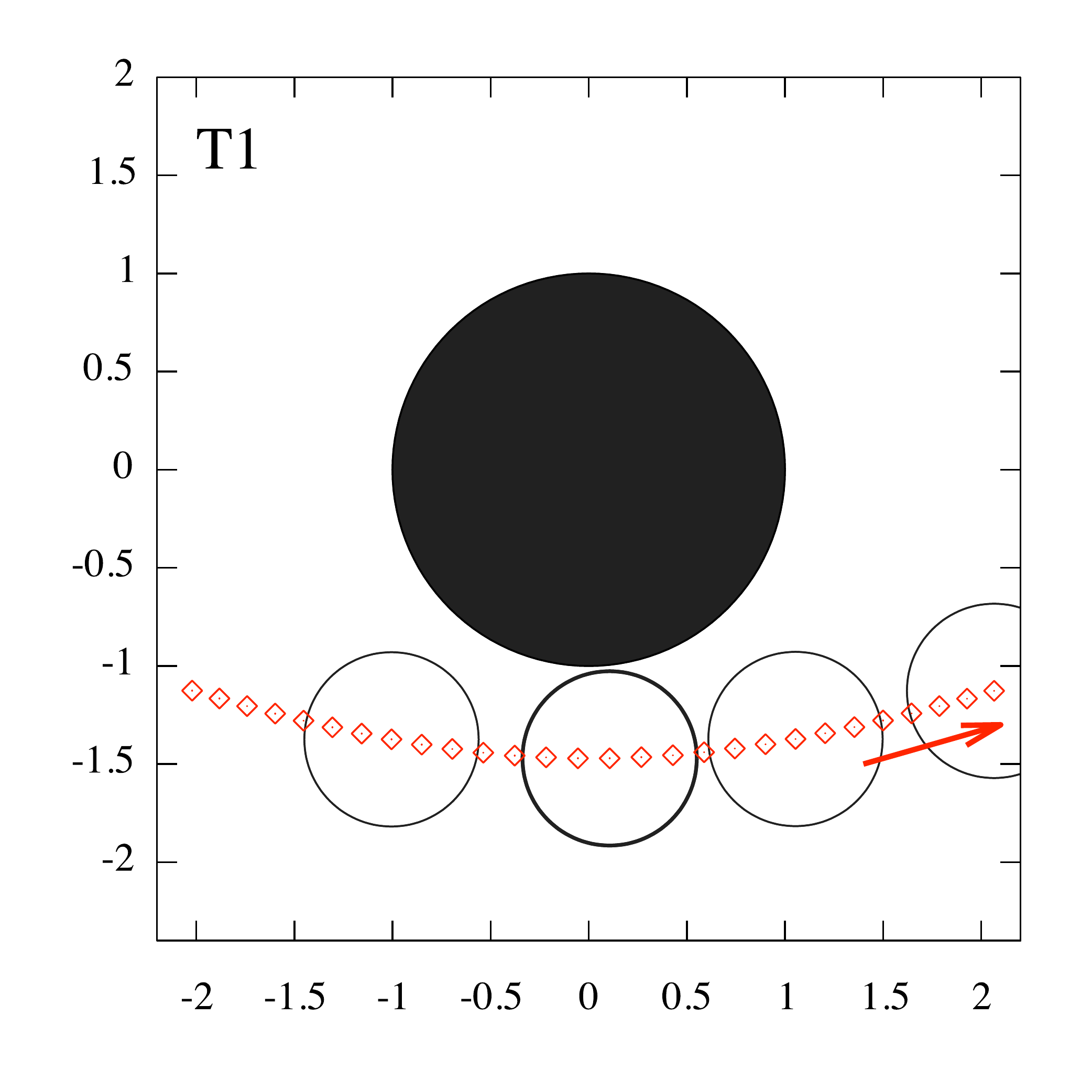} \\
\includegraphics[width=3in]{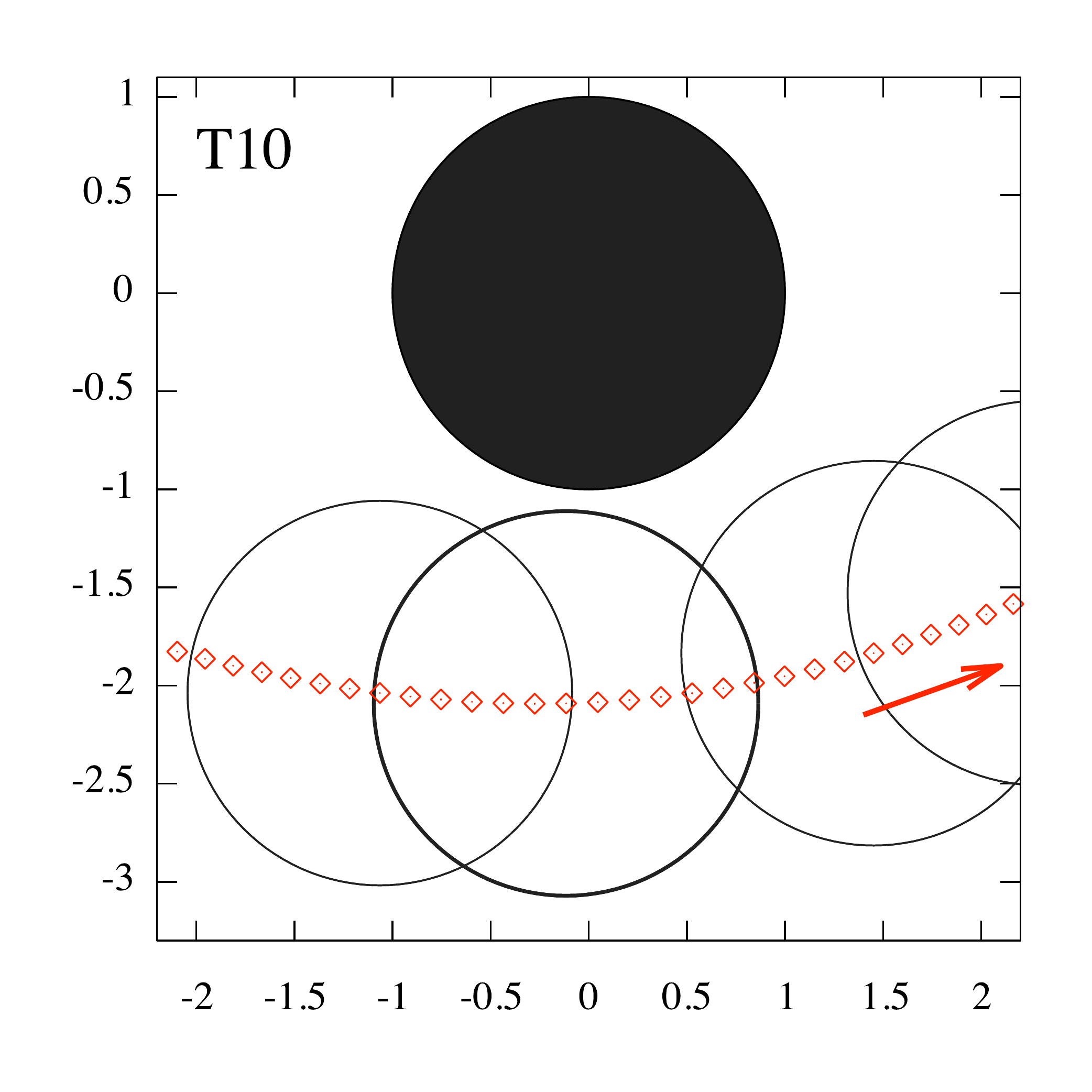} 
\includegraphics[width=3in]{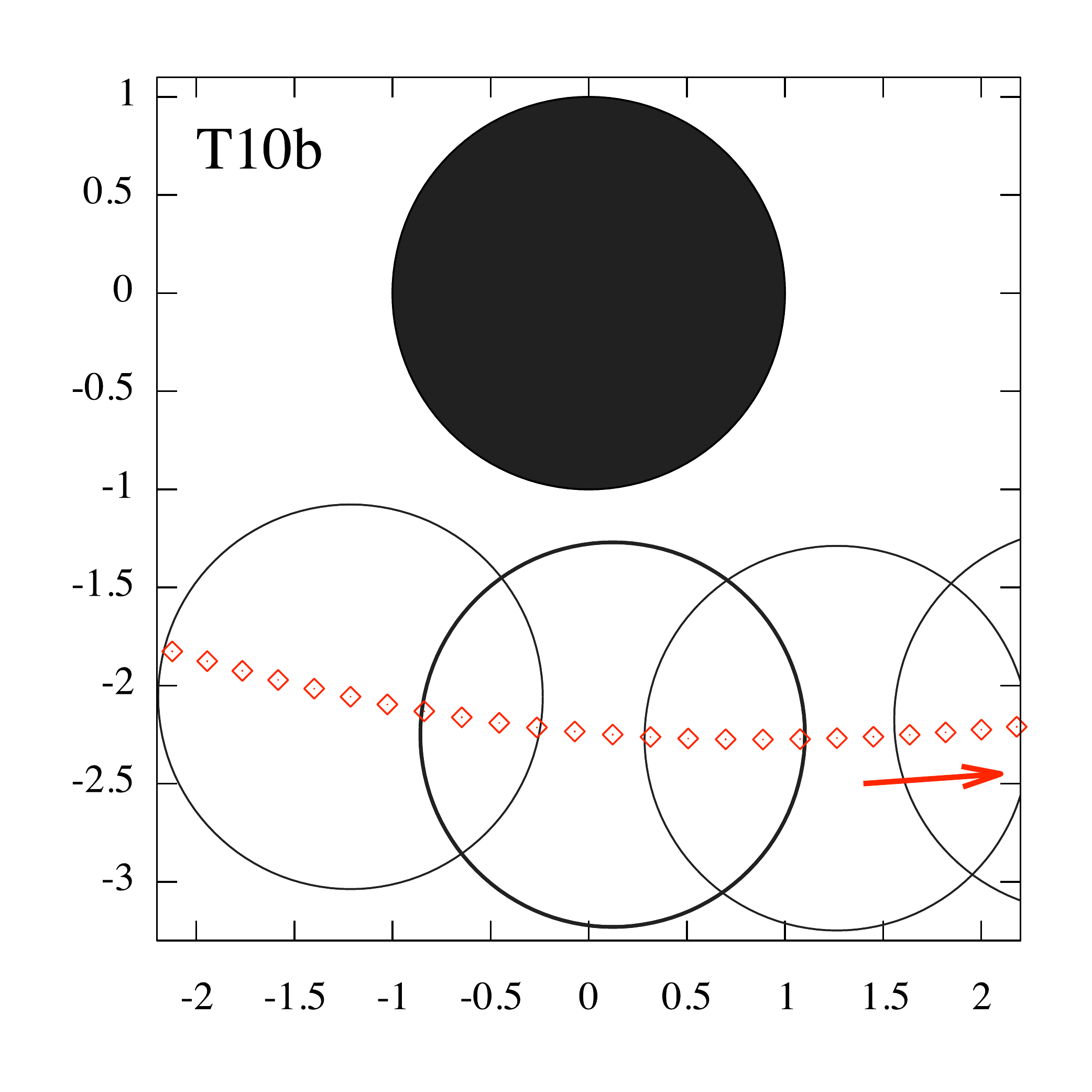} 
\caption{The trajectory of the perturbing body in the orbital plane at different times for each of the tidal encounters
shown in Figure \ref{fig:T5}.  The coordinate system is with respect 
 to the resolved body shown as a solid black circle at the origin.
Red diamonds show the positions of the center of the perturbing body at times separated by 0.1.
The open circles show the perturbing body at the times of snapshots shown in Figure \ref{fig:T5}.
The viewer for these snap shots  is located at negative $y$ and looking upwards in the figures shown here.
The red arrows show the direction of motion of the perturber.
}
\label{fig:orbit}  
\end{figure*}

\begin{figure*}
\includegraphics[width=7.0in]{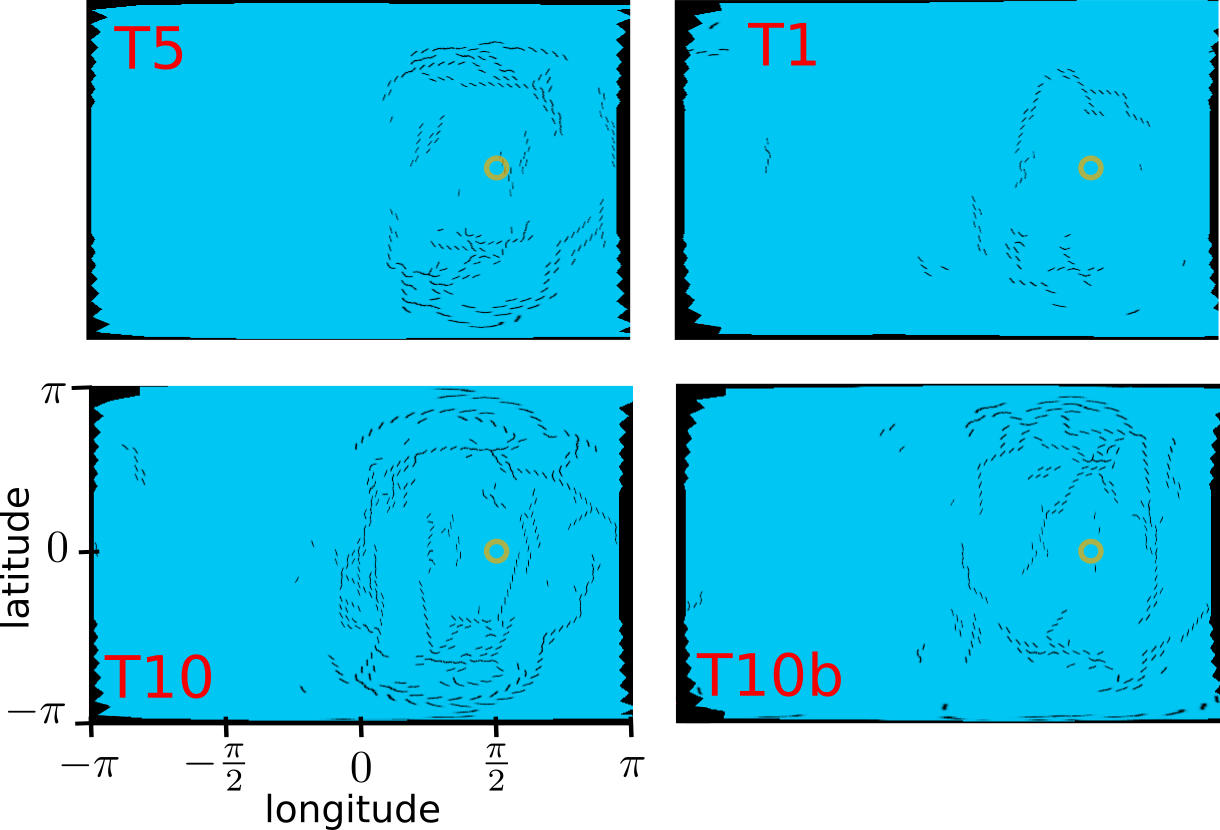} 
\caption{Surface fractures as a function of longitude and latitude (using cylindrical projection) for the same simulations
as shown in Figures \ref{fig:T5}.    Point of closest approach (subsatellite point) is shown as a yellow ring on the middle right
in each panel.
\label{fig:lat}  
}
\end{figure*}

\section{Fractures from Simulated Strong Tidal Encounters}

We illustrate four tidal encounters with body parameters listed in Table \ref{tab:sim_list} under the column
T-series and with encounter parameters listed in Table \ref{tab:sim_qlist}.  
The T5 simulation has perturber $M_2 = 0.5$, half the mass of the primary body.
The T1  simulation  has a  lower strain failure parameter $\epsilon_S = 0.002$
 allowing crustal failure with a lower mass perturber, $M_2 = 0.1$.
The T10, T10b simulations have an equal mass perturbers $M_2=1.0$ and
the T10b encounter is faster than the T10 encounter.

The morphology of the simulated fractures are shown in Figure \ref{fig:T5} using an  orthographic projection, 
and showing an entire  hemisphere. 
In these figures, the
tidal perturber passes from right to left in the plane perpendicular to the line of sight.  
For each simulation, four panels are shown at different
times in the simulation, advancing from the leftmost to the rightmost panel.
Pericenter occurs
with perturber almost directly in front of the primary body and blocking our view of it.   
We render the secondary body with the same density as the primary body and
as nearly transparent so that the surface of the primary body can still be seen.
The trajectories of the perturbing body with respect to our resolved body at different times in the orbit and in the orbital plane
are shown in Figure \ref{fig:orbit} for the same tidal encounter simulations shown in Figure \ref{fig:T5} and listed in
 Table \ref{tab:sim_list}.  
 The resolved body, modeled with the mass-spring network, is shown as a black circle at the origin.
The viewer for the snapshots shown in figure \ref{fig:T5}  is located at 
 at negative $y$  in Figure \ref{fig:orbit} and looking upwards.  Open circles illustrate the locations of the perturbing body
for the snapshots shown in Figure \ref{fig:T5} and has radius set from its mass and assuming that it has the
same density as the resolved body.   Even though the perturber is modeled as a point mass, we show
a radius for the perturber so that the reader has a feeling for the proximity of the two bodies during the tidal encounter.
 If the perturbing body has a similar density as the primary,   these are nearly grazing encounters.

Using the orbital plane to define an equatorial plane for the body, the surfaces after the encounters are shown in Figure \ref{fig:lat}
in a cylindrical projection (horizontal axis corresponding to longitude and vertical axis showing latitude) for
the same simulations.  In these figures with longitude ranging from $-\pi$ to $\pi$ the 
subsatellite point (point of closest approach during the encounter) lies
on the equator at a longitude of $\pi/2$ and on the right hand side.

Figure \ref{fig:T5} shows that simulated crustal fractures extend a large fraction of the body, even for the lowest
mass perturber.   Cracks are oriented both perpendicular and parallel to the orbit path, and are predominantly  present on a single
hemisphere (see Figure \ref{fig:lat}).  Cracks tend to be concentric around the point of closest approach (also called the subsatellite point).
As expected, the faster encounter T10b causes fewer fractures but the fractures seem
to cover the same extent.    
For only the lowest mass perturber are the fractures confined to nearer the subsatellite point.

\begin{figure*}
\begin{tabular}{ll}
\includegraphics[width=2.1in]{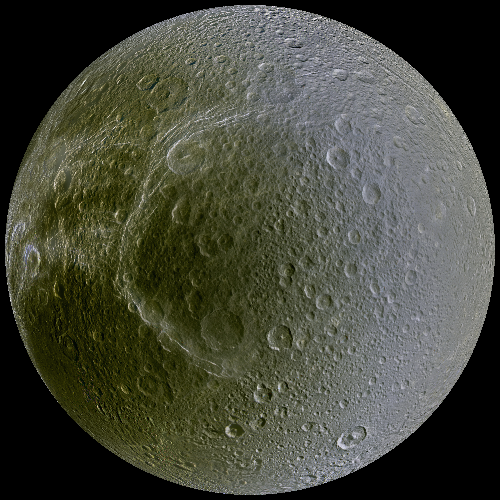} &
\includegraphics[width=2.1in]{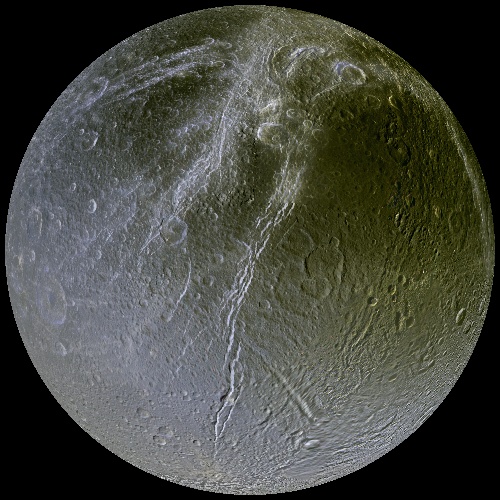} \\
\includegraphics[width=2.1in]{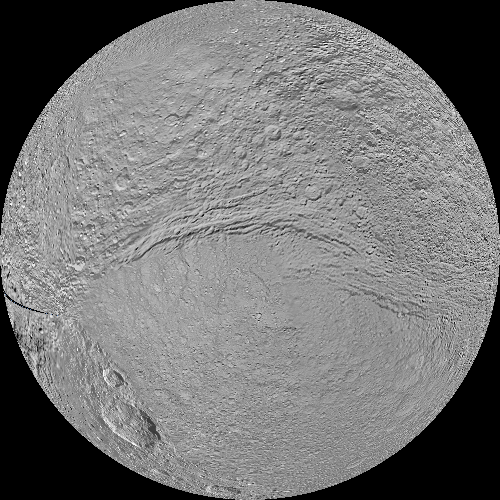} &
\includegraphics[width=2.1in]{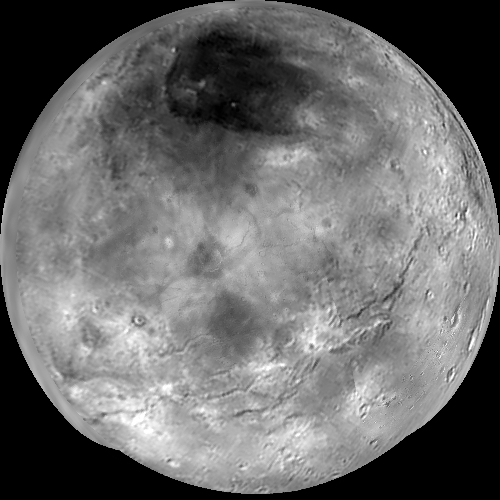}
\end{tabular}
\caption{Images of Dione, Tethys and Charon using an orthographic projection and showing entire hemispheres.  
Top left: Dione using Planetary Image Atlas 18434.  The    
central view point has latitude $16^\circ$, longitude $35^\circ$ and up corresponds to an azimuthal heading of $4^\circ$ from North.
The C shaped feature is Padua Chasmata.
Top Right:  Also Dione but the central point has latitude $-42^\circ$, longitude $131^\circ$, and the heading $9^\circ$.
This image shows Palatine and Eurotas Chasmata.
Bottom Left:  Tethys using   Planetary Image Atlas 11673 and showing the Ithaca Chasma.  The central point has latitude  $-38^\circ$
longitude $18^\circ$, and heading $289^\circ$.
Bottom Right:  Charon using  Planetary Image Atlas 19866.  The central point has latitude  $42^\circ$
longitude $19^\circ$, and heading $8^\circ$.  The large fractures are informally known as Macross and Serenity chasmata.
}
\label{fig:dione}
\end{figure*}

Our simulations produce crude illustrations of surface fractures that we can compare to long chasmata or graben complexes on icy bodies
such as those listed in Table \ref{tab:bodies}.
 In Figure \ref{fig:dione} we show full hemispheres using an orthographic projection 
of Dione, Tethys and Charon created from maps available from the The Jet Propulsion Lab Photojournal 
(see \url{http://photojournal.jpl.nasa.gov/}).
The images shown in Figure \ref{fig:dione} were made using the open-GL display software
\texttt{Sphere-Mapper} (\url{https://github.com/dmgiannel/Sphere-Mapper}), written by one of us (David Giannella),
that takes cylindrical projection cartographic maps and applies them as a texture
to a sphere that can be tilted to any desired angle and rotation angle and viewed using an orthographic projection.
The Dione input image  is the
Planetary Image Atlas 
(PIA) 18434 that is a 
global 3-Color map of Dione (IR-Green-UV) posted 
April 2014.  Its cartographic control and digital mosaic construction are by Dr. Paul Schenk
(LPI, Houston).  The original map has a simple cylindrical map projection at 250m/pixel at equator and is based on 
Cassini ISS images acquired 2004-2014.
The global map of Saturn's moon Tethys (PIA 11673) was created using images taken by NASA's Cassini spacecraft and includes 
new data collected during Cassini's Aug. 14, 2010, flyby (original Image Credit:  NASA/JPL/Space Science Institute).
The map of Charon we used (PIA 19866) is by 
NASA/Johns Hopkins University Applied Physics Laboratory/Southwest Research Institute
and was created from all available resolved images of the surface acquired between July 7-14, 2015, 
at pixel resolutions ranging from 40 kilometers on the anti-Pluto facing hemisphere (left and right sides of the map), 
to 400 meters  per pixel on portions of the Pluto-facing hemisphere. 

The fractures we have simulated (Figures \ref{fig:T5}, \ref{fig:lat}) are of similar extent to those exhibited by the bodies shown in Figure \ref{fig:dione}.
However our simulated bodies often exhibit more than one large fracture, and only Dione has as many chasmata.
Dione has a number of large chasmata, but they are not concentric about a single point as seen on our simulated fractured bodies.
The C shaped feature on Dione known as Padua Chasmata, shown on the top left in Figure \ref{fig:dione}, might originate from a very close
but lower mass encounter with closest approach at the center of the C.
Ithaca Chasma on Tethys, as a single set of features, might be consistent with a fracture caused by a lower mass perturber 
(1/10 of that of Tethys) or more distant encounter with a massive perturber (equal mass).
The long sequence of  chasmata  on Charon forming a great arc
(Macross and Serenity chasmata)\footnote{Names of features on Charon are still informal.} 
might have been caused by  a moderately distant encounter with a very large
object, such as Pluto itself.    Our simulations do show parallel sets of fractures (for example in the T10 simulation)
that might correspond to a series of parallel chasmata such as
Tardis and Nostromo Chasmata that are parallel to Macross and Serenity Chasmata on Charon.

\subsection{Discussion}

Within the  bright terrain on Ganymede is a mosaic of ridges and troughs, termed grooved terrain, exhibiting abundant evidence of extensional strain  (e.g., see \citealt{pappalardo95} and \citealt{collins10} section 6.1.2).
Ganymede has a higher energy density, $e_g \sim 30$ GPa, than the icy bodies we listed in Table 
\ref{tab:bodies}, and this lies in between the Young's modulus of ice (a few GPa) and rocky materials 50-- 100 GPa.
If Ganymede were approximated with an elastic solid with Young's modulus similar to materials in the Earth's lithosphere,   
$E \sim 100$ GPa, then equation \ref{eqn:strain2}
suggests that a near equal mass perturber tidal encounter would crack Ganymede's surface.
Due to Ganymede's liquid metal core and its few hundred km deep           
subsurface saltwater ocean (e.g., \citealt{saur15}), Ganymede would 
deform more strongly to a tidal encounter than a purely elastic body.  
Ganymede is differentiated so a model with a multiple layer interior, comprised of both solids and liquids,
is required to study its tidal response.   

We have taken care in our simulations to relax the body before each encounter and we have tried to 
model a  crust with constant thickness, uniform elasticity and flexural rigidity that is approximately hydrostatically supported.
Even though we have used a moderate number of
particles to resolve shell and interior, the random particle distribution of the interior and associated 
spring network is not even.
There are different numbers of cross springs per shell particle so
some areas of the surface shell are more likely to fail than others and this prevents us from running at low levels
of maximal strain $\epsilon_S$. 
Conversely icy crust on
 moons  is unlikely to have uniform thickness and composition and could be under residual localized stress.  
 Our simulations predict
symmetrical features above and below the orbital plane,  but real bodies are likely to be heterogeneous in terms of crustal thickness, thermal state and pre-existing fractures so they might  preferentially fracture in weaker  regions and
 only on one side.  
Here we have neglected body spin.  However with a nearly parabolic (or slow) encounter with  a rapidly spinning
body (near the breakup spin rate), tidal stresses on the surface would not be symmetrical  
and this too could cause asymmetry in the fracture  distribution.

We found that our simulated bodies did not exhibit fractures for encounters with larger pericenter distances
or higher encounter velocities.    
The escape velocity from Dione is $\sim 0.5$ km/s and this is representative for the escape velocities
for the other icy bodies listed in Table \ref{tab:bodies}.   In comparison, the orbital velocity of Dione  is approximately 10 km/s
and the orbital velocity of Saturn is 9.6 km/s.
This implies
that tidal encounters with asteroids or centaurs would preferentially be at higher relative velocities with 
ratio $V_q/v_c \sim 10$ 
and so higher than the parabolic encounters consider here 
(see equation \ref{eqn:strain2} for the estimated strain dependence on encounter velocity). 
However, encounters with moons that are also orbiting the planet would have much lower relative velocities.
The orbital speed of Charon is only 0.2 km/s so an encounter
between Charon and Pluto would have been in the nearly parabolic regime simulated here.
While a fast encounter from an external object would be isolated, if a close tidal encounter occurs
between two satellites orbiting the same body, then multiple close encounters are likely, each producing
a group of fractures about   a different pericenter locus.

Vibrations are excited during a tidal encounter and vibrational energy dissipated due to the viscoelastic body response 
and surface fracture.
The total energy dissipated should be a small fraction (at most a few percent) 
of the gravitational binding energy  \citep{press77} 
and so is at most a small fraction of the orbital energy.  Only if the encounter is almost exactly parabolic would
this energy loss allow the two bodies to become gravitational bound (in orbit about each other) after the encounter. 

Here we have modeled the perturbing body as a point mass.  If this body is a strong solid and nearly
spherical this is a good approximation.  However if the perturbing body were 
 weak (a low cohesion rubble pile, e.g., \citealt{richardson09}) 
 and lower mass than the primary body then it could be strongly deformed or disrupted 
during then encounter and its tidal field would significantly differ from that of a point mass.

Many large icy moons are believed to have global oceans which decouple the motions of their
 floating shells from their interiors \citep{schubert04,thomas16}  and these 
 could make the tidal response  larger than estimated from elasticity of a solid body alone 
(e.g.,   \citealt{iess12}).  
Conversely,  in our simulations we have neglected the presence of a rocky core and this would have reduced
the tidal response compared to that simulated here using a uniform body.
The maximum strain value for spring
failure $\epsilon_S = 0.003 $ for most of our simulations may be an over estimate for the brittle strength
of ice, so faster and more distant tidal encounters might also be able to cause crustal fractures in icy satellites.
Unfortunately we cannot yet run simulations with lower levels of $\epsilon_S$ because the surfaces
tend to exhibit fractures
even in the absence of a perturber.   We have also been unable to run encounters
with larger mass bodies because in this regime, the shell can be lifted off the body, our surface
triangles collapse and the hinge forces cause the shell to become unstable (and explode).
So far we have only simulated elastic materials and do not have the capability to simulate
simultaneously elastic and liquid materials, though we could increase the 
density and strength in the core by adjusting particle masses and spring constants so we could mimic the behavior
of a rocky core.
 We would like to improve our code to improve its precision and extend the types of materials  that we 
 can simulate.

 We have briefly explored the effect of a non-linear spring force law  on the cross springs by
 multiplying the spring forces (equation \ref{eqn:Fe}) by a strain dependent factor  \citep{clavet05}
\begin{equation}
f(\epsilon) = 
\begin{cases} 
1                                                       & \mbox{if } \epsilon \le 0 \\
\max \left[ \left(1 - \frac{\epsilon}{\epsilon_C} \right), F_{Ck} \right] & \mbox{if }  \epsilon > 0 
\end{cases}
\end{equation}
where $\epsilon_{C}$ sets a strain scale and the spring force is reduced as this scale is approached.
Here  $F_{Ck}$ is a factor that limits the minimum force under extension so that it never drops to zero.
However for the strengths of the bodies we have simulated here, and using the same initial relaxed body,
we saw little difference in simulated fracture morphology when we varied $\epsilon_C$ from 0.02 to 0.04 (and the maximum
strain value was never approached, possibly because the cross springs begin under compression).
Were we to simulate a weaker (so thinner) surface, a reduction in the connection of the surface  to the interior 
would affect simulated fracture extent and morphology.

Lithospheric brittle failure depends on depth, tensile mode, strain and strain rate (see the review by \citealt{burov11}).
Here we have adopted a simplistic maximal strain value for brittle failure.
Crack propagation is only crudely simulated here, and crack morphology is numerically dependent on the nature
of relaxation and so here on the time interval used to identify failed springs in the simulation.   
When this interval  is reduced (and comparing simulations beginning with the same relaxed body), there is some additional surface
failure and in some regions wider cracks can form.  
To meaningfully predict fault lengths, widths and structures we would need to better simulate crack formation, 
propagation, decompression 
and subsequent relaxation.
Wholesale crustal failure in a model with 
more realistic coupling between crust and interior might exhibit fluid flow from subsurface oceans and decompression melting.

We have not seen qualitative differences in fracture morphology if we set rest lengths of springs to their failure length when they fail.
For the strongest perturber (the T10) simulation large regions of the surface were ruptured.  If crack propagation were
more accurately modeled with higher resolution, these might manifest as jumbled terrain or parallel sets of fractures and graben complexes.

Here we have only simulated brittle elastic response.  At the high strain rates of tidal encounters there must be
a depth where the ice is ductile and deforms plastically.   If we more accurately modeled the strength envelope and material
properties as a function of depth, then fractures would open during the encounter
but they would not close all the way afterwards, giving us an estimate for their extension.

We have not modeled viscoelastic response on long time scales.    While we might have identified regions where
fractures originate, our work does not predict depths and widths of resulting chasmata or graben complexes that might result. 
Improved simulations would more accurately model crack propagation, fracture formation and explore the
 subsequent evolution of the resulting geophysical structures, such as
magna and water escape along dilatant cracks and associated partial resurfacing.

\subsubsection{Cracking Mars' Lithosphere}

Using the Young's modulus of a rocky material,  $E_{rock} \sim$ 50-100 GPa,
we notice  that Mars has ratio of 
 gravitational energy density to Young's modulus 
$e_g/E_{rock} \sim $ 2--4, only a factor of  a few higher than $e_g/E_{ice}$
estimated for the icy bodies we have listed in Table \ref{tab:bodies}.
The yield strength envelope as a function of depth for a rocky lithosphere reaches a maximum
at about a few hundred MPa (e.g., Figure 6.35 by  \citealt{watts01}).    The maximum strength
divided by a Young's modulus of 50 GPa (for a rocky material) gives a ratio of order 0.01,
 similar  to the value of maximal strain for uniaxial tensile brittle failure of
ice that we used above as a criterion to mimic fracture in an icy crust.
The ratio $e_g/E$ and the maximum uniaxial strain for crustal brittle failure
are similar for icy moons and Mars, suggesting that 
that the scenario we have explored here can be scaled to Mars.
If Mars experienced a strong (nearly equal mass) and
close tidal encounter with a rocky body, tensile deformation during
the encounter could have fractured  Mars' lithosphere.

Valles Marineris is thought to have been formed by crustal extension \citep{tanaka89}
similar to the formation of rift faults like the East African Rift  \citep{ebinger91}.
Valles Marineris was formed after much of the volcanic
Tharsis bulge or rise was in place (see section 4.5 by \citealt{golombek10}).
Tectonic formation scenarios for Valles Marineris associate its formation
with the inability of Mars' lithosphere to support the large load of the Tharsis bulge itself
\citep{tanaka89,mege96,nimmo05,golombek10, andrewshanna12a, andrewshanna12b, andrewshanna12c}.
Flexural (bending and membrane) stresses in the lithosphere
account for the radial grabens on the topographic rise and concentric
compression wrinkle ridges around its circumference  \citep{banerdt00,golombek10}.
However, a stress model that is azimuthally symmetric about the center of the Tharsis rise  would
predict radial grabens that are of similar length and width  on
opposite sides of the rise.  In contrast, Valles Marineris (south east of the peak) is exceptional -- much
wider and deeper than radial grabens to the north or west of the center of the rise. 

Gravity and topography data indicate that Mars' crustal thickness is bimodal, approximately
30 km in the northern hemisphere and 60 km in the souther hemisphere,
and exceeding 80 km on the Tharsis bulge \citep{neumann04}.
\citet{andrewshanna12b} proposed that Valles Marineris is
located near and aligned with
the buried crustal dichotomy boundary that
bisects the Tharsis bulge.
In his model, the difference in crustal thickness underlying
the Tharsis bulge generates tensile stress directly along
the crustal dichotomy boundary and this accounts for the exceptional width and depth 
of Valles Marineris compared to other radial grabens surrounding the Tharsis bulge.

We have found that a strong tidal encounter can cause 
long extensional fractures that extend a significant fraction of the body's radius.
So a tidal encounter might account for the exceptional depth and length of Valles Marineris.  
However, we expect (see Figure \ref{fig:lat}) 
 fractures in a large ring centered at the point of closest approach.  
If a tidal encounter were responsible for formation of Valles Marineris why
doesn't the valley  extend further, forming a large ring?
We simulated a relaxed, constant thickness, uniform elasticity and flexural rigidity  crustal shell 
but Mars' lithosphere is not uniform thickness and has variations in levels of localized stress.
During a tidal encounter,
Mars' lithosphere could have preferentially fractured along a region of localized stress rather 
than in a circle centering the point of closest approach.
Magma underlying the Tharsis bulge at the time of the tidal encounter may have allowed Mars' crust to be more
easily deformed during the encounter, perhaps even lifted away from the  core.  So
the volcanic activity in the Tharsis province itself could have exacerbated tidal deformation.
A tidal encounter is an intriguing alternate explanation for the extensional stress 
forming Valles Marineris, but a  more sophisticated study would be needed to test it and contrast it
with tectonic models for the formation of Valles Marineris.

\section{Summary and Additional Discussion}

In this paper we have explored tidal encounters with elastic bodies using a mass-spring model
to simulate elasticity within the context of an N-body simulation.  
We have simulated crustal failure using an elastic shell model, with flexural
stiffness for the crust.
Brittle failure is modeled using a maximum strain value for the surface springs, and crack propagation modeled
with a crude relaxation procedure, allowing only a single spring to fail in a specific time interval (a few computational time steps). 
Our simulations illustrate that strong, close tidal encounters can cause crustal failure on icy bodies, confirming
an order of magnitude estimate for the  tidally induced surface strain.  
Following the encounters, simulated crustal shells exhibit long fractures
extending over a large fraction of  a body radius.    Using near parabolic encounters with a nearly equal mass body
perturbing a non-spinning  body,
we find that surface fractures tend to be concentric around the subsatellite point (point of closest approach for the encounter) 
and are restricted
to a single hemisphere.  Tidally induced crustal fractures might provide an explanation for long chasmata and graben complexes
on icy bodies such as Dione, Tethys and Charon.

We have attempted to construct simulations that represent icy bodies.  However, the tidal regime, because it is at high
strain rate, is  different than other geophysical settings (such as studies of  crustal plate flexure).
If some  chasmata are explained by tidal encounters we might be able to 
place constraints on the effective crustal elastic thickness at high strain rate and on the connection between crust
and interior.  Conversely uncertainty in the rheological models  make it difficult for
us to carry out simulations accurate enough to do this comparison.

We have focused here on icy bodies that have old crusts.   Rocky satellites and asteroids such as 
Phobos, Eros, Ida, Gaspra, Epimetheus and Pandora can exhibit long grooves and troughs 
 (see the review by \citealt{thomas10}).   Single tidal encounters could be investigated
as a mechanism for the formation of fractures that are suspected to underly their surface regolith. 
Lastly, strong close tidal encounters might have occurred with large rocky bodies such as Mars, 
and old regions of their surfaces may retain extensional features caused by these encounters.

\vskip 0.1 truein

{\bf Acknowledgements}
We thank 
Maciej Kot,  Moumita Das, Julien Frouard, Stephen Burns, 
Alexander Moore, Melissa Morris, April Russel, Matt Hedman, 
Hanno Rein, Rob French,
Matt Tiscareno, and Shoshanna Cole for encouraging
and helpful discussions. 
This work was in part supported by NASA grant NNX13AI27G.

\end{document}